\documentclass[twocolumn,superscriptaddress,floatfix,preprintnumbers,prd ,nofootinbib,hyperref]{revtex4-2}
\pdfoutput=1
\usepackage[colorlinks=true,breaklinks=true]{hyperref}
\usepackage[normalem]{ulem}
\usepackage[utf8]{inputenc}
\hypersetup{allcolors=[rgb]{0.0 0.0 0.6},linkcolor=[rgb]{0.75 0.05 0.05}}
\usepackage{amsmath,amssymb}
\usepackage{mathtools}
\usepackage{cancel}
\usepackage{tensor}
\usepackage{amsmath, amssymb}

\usepackage[separate-uncertainty=true,multi-part-units=single]{siunitx}
\usepackage{dcolumn}
\usepackage{braket}
\usepackage{multirow}
\usepackage[dvipsnames]{xcolor}
\usepackage{float}
\usepackage{bbold}
\usepackage{letltxmacro}
\LetLtxMacro{\oldcite}{\cite}
\renewcommand{\cite}[1]{\mbox{\oldcite{#1}}}

\newcommand{\dd}{\mathop{\mathrm{d}\!}{}}

\begin{document}
\preprint{DESY-26-146}
\preprint{CERN-TH-2026-274}
\title{Stars Can Quench Superradiance Around Supermassive Black Holes}

\author{Yu Cheng}
\email{chengyu059@uchicago.edu}
\affiliation{Department of Physics, Enrico Fermi Institute, Leinweber Institute for Theoretical Physics,
Kavli Institute for Cosmological Physics, University of Chicago, Chicago, IL 60637, USA}
\affiliation{Department of Physics, Korea Advanced Institute of Science and Technology (KAIST), Daejeon 34141, Korea}

\author{Andrea Caputo}\email{andrea.caputo@cern.ch}
\affiliation{Department of Theoretical Physics, CERN, Esplanade des Particules 1, P.O. Box 1211, Geneva 23, Switzerland}
\affiliation{School of Natural Sciences, Institute for Advanced Study, Princeton, NJ 08540, USA}
\affiliation{Dipartimento di Fisica, ``Sapienza'' Universit\`a di Roma \& Sezione INFN Roma1, Piazzale Aldo Moro
5, 00185, Roma, Italy}

\author{Giovanni Maria Tomaselli}\email{tomaselli@ias.edu}
\affiliation{School of Natural Sciences, Institute for Advanced Study, Princeton, NJ 08540, USA}

\author{Samuel J. Witte}\email{Samuel.Witte@physics.ox.ac.uk}
\affiliation{Rudolf Peierls Centre for Theoretical Physics, University of Oxford, UK}
\affiliation{Deutsches Elektronen-Synchrotron DESY, Notkestraße 85, 22607 Hamburg, Germany}
\affiliation{
II. Institute of Theoretical Physics, Universität Hamburg, 22761, Hamburg, Germany}

\begin{abstract}

Light scalar fields trigger an instability around rotating black holes which drives the rapid growth of high-density bound states on astrophysically short timescales---a phenomenon known as black hole superradiance. This process is usually tacitly assumed to happen independent of the local environment. Around supermassive black holes, however, orbiting stars and compact objects act as a perturbation that drains energy from the cloud, potentially preventing its formation altogether. In this work, we quantify under what conditions this happens, focusing on bound-free state transitions. We first compute the rate of energy dissipation due to gravitational interactions, finding that the superradiance growth rate almost always dominates the environmental depletion. The two rates are, however, close enough that the conclusion could be overturned once bound-bound transitions are included. Also, the environment dissipates the cloud faster than gravitational-wave emission, implying that stellar environments must be accounted for in a variety of indirect search strategies. We then investigate the impact of the scalar's quadratic couplings to nucleons and electromagnetism, showing that these interactions can greatly enhance energy dissipation, preventing the cloud's formation. This work highlights the importance of accounting for the local environment in studies of superradiance around intermediate and supermassive black holes.

\end{abstract}

\maketitle

\section{Introduction}

Black hole superradiance is a phenomenon in which light bosonic fields experience an instability around rotating black holes~\cite{ZelDovich1971,ZelDovich1972,Starobinsky:1973aij,Brito:2015oca}, allowing these particles to rapidly extract energy and angular momentum from the black hole (in a manner analogous to the Penrose process~\cite{1971NPhS..229..177P}) on short time-scales, leading in turn to the appearance of high-density quasi-bound states of the bosonic field. This process is typically efficient when the Compton wavelength of the scalar is comparable to, but slightly larger than, the black hole horizon. From an observational perspective this process has a number of interesting implications, including an incompatibility of rapidly spinning black holes and light bosonic fields (where the mass scale of the constrained boson is approximately set by the mass of the black hole)~\cite{Arvanitaki:2010sy,Arvanitaki:2014wva,Hoof:2024quk,Witte:2024drg,Caputo:2025oap,Aswathi:2025nxa}, the production of gravitational wave lines~\cite{Tsukada:2018mbp,Palomba:2019vxe,Ng:2020jqd,LIGOScientific:2021rnv,Yuan:2022bem,Ellis:2026gwt}, the modification of binary inspirals~\cite{Arvanitaki:2010sy,Arvanitaki:2014wva,Arvanitaki:2016qwi,Ferreira:2017pth,Hannuksela:2018izj,Zhang:2019eid,Palomba:2019vxe,Baumann:2018vus,Tsukada:2020lgt,Baumann:2019ztm,Baumann:2021fkf,Baumann:2022pkl,Tong:2022bbl,Siemonsen:2022yyf,Fan:2023jjj,Brito:2023pyl,Tomaselli:2023ysb,Boskovic:2024fga,Tomaselli:2024bdd,Tomaselli:2024dbw,May:2024npn,Boskovic:2025ixx,Roy:2025qaa,Xu:2026cky}, and indirect effects coming from the existence of the high-density field configurations~\cite{GRAVITY:2023cjt,GRAVITY:2024tth,Tomaselli:2025zdo}. In the context of axions, superradiance is a leading mechanism for establishing constraints on axions with masses $\mu \sim 10^{-12}\, {\rm eV}$ and $\mu \sim 10^{-19}\, {\rm eV}$, where the former is set by the observations of  black holes with $\mathcal{O}(10) \, M_\odot$ and the latter being set by supermassive black holes~\cite{Hoof:2024quk,Witte:2024drg,Caputo:2025oap,Aswathi:2025nxa}. 

The instability timescale of axion superradiance around stellar mass-scale black holes can be as short as a few minutes\footnote{Note that for spin 1 and 2 particles, these characteristic timescales are in fact much shorter.}, making it incredibly challenging for astrophysical processes to alter the early-stage dynamical evolution of these systems. The same is not true, however, for supermassive black holes, where so-called `spin-down' timescales are typically on the order of kyrs, or longer. These objects also sit in the central regions of galaxies, and are thus also inherently situated in more complex astrophysical environments. For example, the black hole sitting in the center of the Milky Way (Sagittarius A$^*$) has a high stellar density environment, with stellar pericenters that pass close to the black hole itself. This raises a very natural question about whether environmental effects in these systems can alter the growth or decay of superradiant clouds (note that the decay rate of the cloud plays a crucial role in setting the sensitivity of dynamical tests and gravitational wave searches for fully grown superradiant systems).

In this work, we investigate how stars in the local environment of a supermassive black hole can perturb the evolution of scalar superradiance away from the naive, idealized limit. In particular, we focus on a number of distinct processes, including: gravitational dynamical friction (a purely gravitational process which assumes nothing about the interactions of these particles), a quadratic coupling to nucleons, and a quadratic coupling to electromagnetism. In each case, we analyze the net energy dissipation from bound-free transitions induced by the local stellar population, highlighting the impact of astrophysical uncertainties associated with the stellar mass distribution, the distributions of orbital eccentricities, etc. Let us emphasize that resonant bound-bound transitions can lead to a much greater rate of energy loss when resonantly driven by particular orbits, and off-resonant bound-bound transitions may contribute comparably to the bound-free contribution. As such, the rate of energy dissipation derived here should be considered as a lower limit to the true contribution within a particular model.

Our results indicate that purely gravitational dynamics of the stellar population are unlikely to generically suppress superradiant growth, although in some cases the rate of energy dissipation becomes comparable to it. Nevertheless, gravitational dynamics are expected to strongly modify the decay rate of the superradiant cloud across a large region of parameter space.  When quadratic interactions are present, the rate of energy dissipation can be enhanced relative to the gravitational scenario by many orders of magnitude, and can in many cases prohibit the growth of superradiant clouds. Our results suggest that caution must be taken when studying superradiance around supermassive black holes, as assumptions on the unknown stellar dynamics or the interactions of the field can strongly impact the growth and evolution of these systems.  

This manuscript is organized as follows. We begin in Sec.~\ref{sec:superradiance} by reviewing various fundamental aspects of black hole superradiance, highlighting the relevant scales of the problem. We describe the stellar population models in Sec.~\ref{sec:stellar-model}, and the calculation of bound-free transitions in Sec.~\ref{sec:ionization}. In Sec.~\ref{sec:gravity}, we derive the energy loss rate induced by gravitational dynamical friction of a single star. We then generalize this result to stellar populations, and compare the collective effect to the timescales which govern superradiant evolution. In Sec.~\ref{sec:couplings}, we compute the energy loss rates induced by generic, slowly varying, quadratic potentials, and apply our results to both the quadratic nucleon and quadratic electromagnetic couplings.  We conclude in Sec.~\ref{sec:conclusions}.

\begin{figure}[!t]
    \centering
    \includegraphics[width=0.468\textwidth]{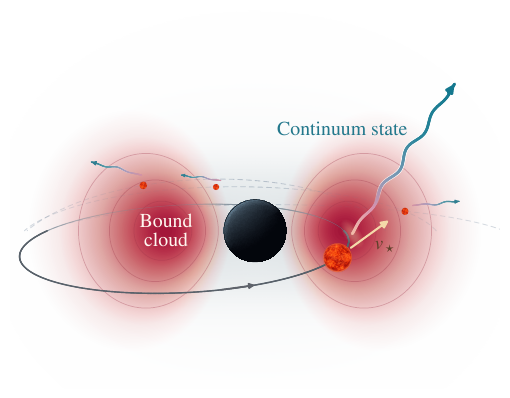}
    \caption{Rough sketch of the process studied in this work. Here, a superradiant cloud forms around a central massive black hole. Stars orbiting the black hole and passing through the superradiant cloud can induce bound-free transitions, modifying the superradiant growth rate or accelerating the decay of the cloud itself. 
    }
    \label{fig:sketch}
\end{figure}
%

\section{Black Hole Superradiance}
\label{sec:superradiance}

In this section, we briefly review various aspects of BH superradiance, highlighting in particular the growth and evolutionary timescales, and the structure of the quasi-bound states. Further details can be found, e.g., in Refs.~\cite{Brito:2015oca,Arvanitaki:2010sy,Baumann:2019eav}.

\subsection{Quasi-bound states}

Let us start by considering an isolated rotating BH in the presence of a non-interacting scalar field $\Phi$. The field dynamics can be obtained by looking for quasi-bound state solutions of the Klein-Gordon equation in Kerr background. The problem can be approached analytically via a matched asymptotic expansion, where the equation is initially solved in the near- and far-field limits using the corresponding boundary conditions, and then the two solutions are matched in an overlapping intermediate region where both solutions remain valid.

The strategy described above is a perturbative calculation whose control parameter is the ``gravitational fine structure'' constant, $\alpha\equiv GM\mu\ll1$, where $M$ is the BH mass and $\mu$ is the scalar mass. The discrete spectrum of quasi-bound states is composed of states labeled by ``quantum'' numbers $\ket{n\ell m}$ (analogous to the hydrogenic orbitals in quantum mechanics) with complex frequencies, $\omega=\omega_r+i\,\omega_i$. The leading order behavior of this perturbative matching scheme yields real and imaginary frequencies given by
\begin{align}
    \omega_r &= \mu \bigg(1 - \frac{\alpha^2}{2 n^2} - \bigg[\frac{1}{8 n} + \frac{6}{2 \ell + 1} - \frac{2}{n} \bigg] \frac{\alpha^4}{n^3}+\ldots \bigg)\,, \\[4pt]
    \label{eqn:detweiler}
    \omega_i &= 2 \tilde{r}_+\, C_{n \ell} \, g_{\ell m} \, (m \Omega_+ - \omega_r) \, \alpha^{4 \ell + 5}\,, 
\end{align}
where $\Omega_+=\tilde a/(2M(1+\sqrt{1-\tilde a^2}))$, $\tilde a$ is the dimensionless BH spin, and
\begin{align}
    C_{n \ell} &\equiv \frac{2^{4 \ell + 1} (n + \ell)!}{n^{2 \ell+4} (n-\ell-1)!} \left[\frac{\ell!}{(2 \ell)! (2 \ell+1)!} \right]^2\,, \\
    g_{\ell m} &\equiv \prod_{k=1}^\ell \left(k^2 (1-\tilde{a}^2) + (\tilde{a} m - 2 \tilde{r}_+ \alpha)^2 \right) \, .
\end{align}
While the analytic forms are useful for understanding the functional scaling, the approximation to $\omega_i$ (which controls the e-folding growth timescale of the superradiant cloud) quickly becomes inaccurate for the fastest growing states at $\alpha\gtrsim0.1$; we therefore employ Leaver's continued fraction method \cite{Leaver:1985ax,Dolan:2007mj,Baumann:2019eav} throughout this work to compute the relativistic energies where relevant.

The eigenfunctions can be decomposed as $\Phi=\sqrt{N/(2\mu)}R(r)S(\theta)e^{im\phi}e^{-i\omega t}$, where $N$ is the occupancy number, $S(\theta)$ is a spin-weighted spheroidal harmonic, and 
\begin{equation}
\begin{split}
    R(r)={}&\sqrt{\bigg(\frac{2 \mu \alpha}{n}\bigg)^3 \frac{(n-\ell-1) !}{2 n(n+\ell) !}}\,\biggl(\frac{2 \alpha \mu r}{n}\biggr)^{\ell}\\
    &\times\exp\biggl(-\frac{\mu \alpha r}{n}\biggr)\, L_{n-\ell-1}^{2 \ell+1}\biggl(\frac{2 \mu \alpha r}{n}\biggr)\,,
\end{split}
\label{eq:ContinueState}
\end{equation}
where $L_{n-\ell-1}^{2 \ell+1}$ is a generalized Laguerre polynomial. The resulting density profiles of the three fastest-growing states ($\ket{211}$, $\ket{322}$, and $\ket{433}$) are, in the non-relativistic limit,
\begin{align}
& \rho_{211}(\mathbf{r})
=
\frac{M_{\mathrm{c}}}{64 \pi r_{\mathrm{c}}^3}\left(\frac{r}{r_{\mathrm{c}}}\right)^2
e^{-r / r_{\mathrm{c}}} \sin ^2 \theta, \\
& \rho_{322}(\mathbf{r})
=
\frac{M_{\mathrm{c}}}{26244 \pi r_{\mathrm{c}}^3}
\left(\frac{r}{r_{\mathrm{c}}}\right)^4 
e^{-2 r /\left(3 r_{\mathrm{c}}\right)} \sin ^4 \theta,\\
& \rho_{433}(\mathbf{r})
=
\frac{M_{\mathrm{c}}}{37748736 \pi r_{\mathrm{c}}^3}
\left(\frac{r}{r_{\mathrm{c}}}\right)^6 
e^{- r /\left(2 r_{\mathrm{c}}\right)} \sin ^6 \theta.
\label{eq:threestatesrho}
\end{align}
where $M_c$ is the cloud's mass, $r_c \equiv GM/\alpha^2$ is the characteristic cloud radius, and $r$ and $\theta$ are polar coordinates with respect to the BH spin direction. Note that we use the non-relativistic wave functions to compute the rate of energy dissipation throughout this work -- as we will show below, the effects studied here are most relevant for smaller values of $\alpha$, and thus don't expect this to introduce a significant error in our comparison.

\subsection{Cloud growth and saturation}

Superradiant states are those quasi-bound states for which $\omega_i>0$, which corresponds to the condition that $\omega_r < m \Omega_+$. The occupancy number (and thus the total mass stored in the superradiant cloud $M_c \simeq \mu N$) grows with an $e$-folding timescale $\tau_{\text{sr}}\equiv1/(2\omega_i)$; in general, this growth proceeds uninterrupted until an order one fraction of the black hole spin has been extracted, and the superradiant condition listed above has been saturated. Generating a fully grown cloud from a single quantum fluctuation requires around $\sim 180$ e-folds of growth, which for Sagittarius A$^*$ roughly corresponds to a timescale of 
$\mathcal{O}({\rm Myrs})$ for the $\ket{211}$ state with $\alpha \approx 0.1$.

The values of $\alpha$ for which a given state grows depend on the BH spin. The threshold spin can be found from the superradiance condition,
\begin{equation}
    \tilde a_{\text{th}}=\frac{4GM\omega_r/m}{1+4(GM\omega_r/m)^2}\,.
\end{equation}
The equation can be inverted and solved for $\alpha$. In this work we consider, for illustrative purposes, a spin of $\tilde a=0.9$ -- while this choice is somewhat arbitrary, it allows us to provide a representative case study. For this choice of spin, the threshold values of $\alpha$ for the three fastest-growing states are
\begin{equation}
\alpha^{211}_{\text{th}}=0.317\,,\quad\;\;\alpha^{322}_{\text{th}}=0.641\,,\quad\;\;\alpha^{433}_{\text{th}}=0.969\,.
\end{equation}

Given an initial spin $\tilde a_i$, the final cloud mass can be found from angular momentum conservation,
\begin{equation}
M_c=M\,\frac{GM\omega_r}{m}(\tilde a_i-\tilde a_{\text{th}})\,,
\end{equation}
and can be as large as $\sim10\%$ of the BH mass for the $\ket{211}$ state and a near maximal spin BH.

\subsection{Cloud decay and self-interactions}

There are two effects that can alter the long-term growth and evolution of a superradiant cloud, and which require additional discussion. First, after saturation and spin-down, the cloud slowly loses energy through the boson cloud annihilations into GW. 
This depletion is much slower than the growth of the corresponding state,
and sets the effective lifetime of the saturated cloud. Note that many searches for superradiant clouds implicitly assume a fully grown cloud -- if a cloud decays quickly, this assumption (and thus the search itself) is invalid. We define the GW depletion timescale for various states and black holes in Appendix~\ref{app:GWdepletion}. We also show in the Appendix~\ref{app:GWdepletion}
that energy dissipation from the environmental losses studied here can, in many cases, dominate over GW emission, suggesting the typical densities of saturated superradiant clouds around large black holes have likely been overestimated.

Another important factor which can strongly influence the growth and evolution of superradiant clouds is the presence of self-interactions. 
In the case of axions, for example, the discrete  shift symmetry of the axion guarantees the presence of quartic interactions of the form $V(\Phi)\supset\mu^2\Phi^4/f_a^2$ (note that similar conclusions hold for other light scalars whose mass is protected by an approximate symmetry); these interactions quickly drive the superradiant system to quasi-equilibria configurations, which delay the spin down of the black hole and result in clouds with suppressed occupation numbers~\cite{Baryakhtar:2020gao,Witte:2024drg}. This statement is true even for exceptionally large decay constants. In this work, we focus specifically on the non-interacting limit -- it would be interesting, however, to study the energy dissipation in the case of axions, as superradiant axion clouds can extend to much larger spatial scales and have a more complex structure than non-interacting scalar fields. We leave such a study to future work.

\section{Stellar population model}
\label{sec:stellar-model}

To determine the impact of environmental effects on the cloud's growth and decay, 
one needs to model the population of objects living in the neighborhood of the supermassive BH. Needless to say, this population is not precisely determined for any system of interest, and thus carries significant astrophysical uncertainties. 
In order to illustrate the impact of various modeling assumptions (e.g. including a realistic stellar mass distribution, or dynamical effects on mass separation), we present our results using three distinct models, which we discuss in detail below. Note that the models should not be interpreted on equal footing -- these are intended to highlight the quantitative effect of including distinct physical inputs, and future work will need to be done on a system-by-system basis in order to better understand how to account for realistic environmental effects.

\subsection{Single-mass cusp}

The simplest model consists of a power-law cusp, in which all stars  have one solar mass, $M_\star = 1\,M_\odot$. Defining the influence radius $r_h$ as the distance enclosing a stellar mass equal to the BH mass $M$, the number of stars with semi-major axis less than $a$ is (see, e.g., \cite{Gurkan:2007bj})
\begin{equation}
    N_{\rm cusp}(<a)=\frac{M}{M_\star}\bigg(\frac{a}{r_h} \bigg)^{3 - \gamma}\,.
\end{equation}
We further assume that the cusp is spherical and isotropic, with a thermal eccentricity distribution, $p(e)=2e$. 
For the standard Bahcall--Wolf profile, the slope is $\gamma=7/4$ and for the Milky Way, the fiducial parameters are $M=\num{4e6}M_\odot$, $r_h=\SI{2}{pc}$, and $\gamma=1.4$~\cite{1999ApJ...527..835A,Genzel:2003cn,Alexander:2005jz,Gurkan:2007bj}. 
These numbers can be generalized to other galaxies by estimating $r_h\simeq GM/\sigma^2$, and relating the velocity dispersion $\sigma$ to the BH mass $M$ through the $M$--$\sigma$ relation~\cite{McConnell:2012hz}. 
This leads to
\begin{equation}
    r_h(M)\simeq 3.16\,{\rm pc}
    \left(\frac{M}{10^7\,M_\odot}\right)^{0.645}\,.
    \label{eq:rh_scaling}
\end{equation}

\subsection{Segregated continuous stellar-mass population}

Next we consider the strong-segregation continuous stellar-mass profile proposed in \cite{Linial:2022big}, described by an isotropic distribution function $f(x,M_\star)$,
\begin{equation}
f(x,M_\star)
\propto M_\star^{-(\gamma+15)/5}
\exp\Big[-M_\star A x^{5/(4\gamma+10)}\Big]\,,
\label{eq:fdistributionP}
\end{equation}
where $x \equiv GM/r-v_\star^2/2$ is the specific binding energy of a Keplerian orbit, $M_\star$ is the stellar mass, and $\gamma$ is a power-law index.
The distribution function is normalized such that $\big[\int \dd^3r\dd^3v\,f(x,M_\star)\big]\dd M_\star$ gives the number of objects with masses in the interval $[M_\star,M_\star+\dd M_\star]$. 
The constant $A$ appearing in Eq.~\eqref{eq:fdistributionP} is chosen such that
\begin{equation}
    \frac{\partial\log f}{\partial\log x}(\sigma^2,M_\star^{\rm min})=1\,,
\end{equation}
where $\sigma$ is the velocity dispersion at the influence radius and $M_\star^{\min}$ is the minimum stellar mass in the population.

As a concrete example, we take $\gamma=-5$, in which case the distribution function is
\begin{equation}
\begin{aligned}
    f(x,M_\star)
    &=0.26\,
    \frac{\sigma^3}{G^3 M^2 (M_\star^{\min})^2}
    \left(\frac{M_\star}{M_\star^{\min}}\right)^{-2} \\
    &\quad\times
    \exp\bigg[-2\frac{M_\star}{M_\star^{\min}}
    \left(\frac{x}{\sigma^2}\right)^{-1/2}\bigg]\,,
\end{aligned}
\label{eq:final_f}
\end{equation}
where the coefficient $0.26$ is obtained by requiring that the total stellar mass in the region $x \ge \sigma^2$ equals the BH mass $M$.
Assuming Keplerian orbits, a thermal eccentricity distribution, and the $M$--$\sigma$ relation, the differential number distribution reads
\begin{equation}
\begin{aligned}
    \frac{d^3 N}{d a\,d M_\star\,d e}
    &= 2.1\,\pi^3\,e\,
    \frac{\sigma^3 a^{1/2}}
    {G^{3/2} M^{1/2} \left(M_\star^{\min}\right)^2} 
    \left( \frac{M_\star}{M_\star^{\min}} \right)^{-2} \\
    &\quad \times
    \exp\bigg[
    -2\frac{M_\star}{M_\star^{\min}}
    \bigg(\frac{2a}{r_h}\bigg)^{1/2}
    \bigg].
\end{aligned}
\label{eq:d3N_da_dm_de}
\end{equation}

\subsection{Multiple populations}

We also consider a model with multiple subpopulations. For the Milky Way galactic center, we adopt the ``Fiducial$\times10$'' model from Ref.~\cite{Generozov:2018niv}, and use simple broken-power-law fits to the number density for practical calculations.
The three populations are low-mass stars, neutron stars, and stellar-mass black holes, with representative masses
$ m_\star=0.3\,M_\odot$, $m_{\rm NS}=1.5\,M_\odot$, and $m_{\rm BH}=10\,M_\odot$, respectively. 

For the number density of low-mass stars, we use
\begin{equation}
n_\star(r)=
\begin{cases}
3.19\times10^5
\bigg(\dfrac{r}{1\,{\rm pc}}\bigg)^{-1.19}\,{\rm pc}^{-3},
& r \le 5 \,{\rm pc},\\[0.8em]
2.54\times10^7
\bigg(\dfrac{r}{1\,{\rm pc}}\bigg)^{-3.91}
\,{\rm pc}^{-3},
& r > 5\,{\rm pc} 
\end{cases} 
\end{equation}
%
For neutron stars, we adopt the same fitting function as in Refs.~\cite{Leane:2021ihh,Foster:2022fxn},
\begin{equation}
n_{\rm NS}(r)=
\begin{cases}
5.98\times10^3
\bigg(\dfrac{r}{1\,{\rm pc}}\bigg)^{-1.70}
\,{\rm pc}^{-3},
& r \leq 2\,{\rm pc},\\[0.8em]
2.08\times10^4
\bigg(\dfrac{r}{1\,{\rm pc}}\bigg)^{-3.50}
\,{\rm pc}^{-3},
&r > 2\,{\rm pc}.
\end{cases}
\end{equation}
For stellar-mass BHs, we also use a two-region power law fit to the same ``Fiducial$\times10$'' profile,
\begin{equation}
n_{\rm BH}(r)=
\begin{cases}
3.34\times10^3
\bigg(\dfrac{r}{1\,{\rm pc}}\bigg)^{-1.99}
\,{\rm pc}^{-3},
& r\le2.5\,{\rm pc},\\[0.8em]
5.78\times10^4
\bigg(\dfrac{r}{1\,{\rm pc}}\bigg)^{-5.10}
\,{\rm pc}^{-3},
&r > 2.5\,{\rm pc}.
\end{cases}
\end{equation}
Because the input model specifies radial number densities rather than a joint distribution in semimajor axis and
eccentricity, we approximate the orbits of all three components as circular.

For comparison with the stellar population models, we select
five S-stars, S2, S29, S38, S55, and S301, with the smallest pericenter distances.
The semimajor axes and eccentricities of the first four are taken from~\cite{GRAVITY:2024tth,Tomaselli:2025zdo}. 
For S301, we adopt the best-fit orbital parameters in~\cite{Dayem:2026ktt}, with 
$a=3.331 \times 10^{-3} \mathrm{pc}$, $e=0.9832$ and the orbital period of $8.68\,$yr.
For the stellar properties, we adopt the following benchmarks, 
with mean density $\bar\rho_\star = 3M_\star/(4\pi R_\star^3)$. 
For S2, we use $M_\star=13.6M_\odot$ and $R_\star=5.53R_\odot$, 
giving $\bar\rho_\star\simeq0.113\,{\rm g\,cm}^{-3}$~\cite{2017ApJ...847..120H}. 
For S29 and S55, we adopt a common main-sequence benchmark with $M_\star=5M_\odot$ and $R_\star=3R_\odot$, 
giving $\bar\rho_\star\simeq 0.26\,{\rm g\,cm}^{-3}$. 
For S38, we instead use a warm giant benchmark with $M_\star=1.3M_\odot$ and $R_\star=5R_\odot$, 
giving $\bar\rho_\star\simeq1.5\times10^{-2}\,{\rm g\,cm}^{-3}$. 
The parameters adopted for S29, S55, and S38 are benchmark choices motivated by stellar evolution estimates, 
rather than individual measurements. 
For S301, we take $M_\star=1.5M_\odot$ and $R_\star=1.4R_\odot$, 
giving $\bar\rho_\star\simeq0.77\,{\rm g\,cm}^{-3}$,
following the benchmark in Ref.~\cite{Dayem:2026ktt}.

\subsection{Stellar radii and inner density profiles}

As we will highlight below, the stellar radii $R_\star$ and density alter the effective scattering cross section and energy dissipation rates. 
For each of the three population models presented above, we use the \texttt{MESA} package~\cite{Paxton:2010ji,Paxton:2013pj} to compute stellar structure profiles in the mass range $0.1M_\odot \le M_\star \le 100M_\odot$. We assume that stars with $M_\star<10M_\odot$ are middle age main sequence, while those with $M_\star>10M_\odot$ are red supergiants. For solar-mass scale stars, we adopt the standard solar model from~\cite{Vinyoles:2016djt}. 
Let us note that in the case of gravitational effects, the drag force depends only on the stellar mass, and thus our conclusions are effectively insensitive to the stellar profile.
For quadratic interactions, these assumptions  may yield slight shifts in the effective potential (as a result of shifts in the stellar radius and density), but we expect these effects to be small relative to the uncertainties in the stellar populations themselves.

\section{Ionization Formalism for
Environmental Cloud Depletion}
\label{sec:ionization}

An orbiting star sources a periodic perturbation of the gravitational atom. 
The cloud depletion process can therefore be described as a transition from the initially occupied quasi-bound state to continuum states. 
In this section, we review the ionization formalism of Ref.~\cite{Tomaselli:2023ysb}.
The local dynamical friction picture~\cite{Hui:2016ltb} also provides an intuitive physical interpretation of the depletion process. 
However, this approach does not resolve the momentum of the final continuum state,
and is therefore mainly suitable for estimating the cloud mass loss for orbits with small eccentricities.

\subsection{Bound to Continuum Transitions}

Consider a cloud initially in the state $|b \rangle\equiv |n_b \ell_bm_b\rangle$ with corresponding wave function
\begin{equation}
  \psi_{n_b \ell_b m_b}(t,{\bf r})
  =
  R_{n_b \ell_b}(r)
  Y_{\ell_b m_b}(\theta,\phi)e^{-i\epsilon_{b} t},
\end{equation}
where $ \epsilon_{b} = -{\mu \alpha^2}/{2 n_b^2}$ is the bound state energy, and the expressions of the radial functions are given in Eq.~\eqref{eq:ContinueState}.

Let $V(t,{\bf r})$ denote the perturbation sourced by an orbiting star. The initial bound state can then be ionized into a continuum state $|k;\ell m\rangle$, whose wave function is given by
\begin{equation}
  \psi_{k;\ell m}(t,{\bf r})
  =
  R_{k;\ell}(r)Y_{\ell m}(\theta,\phi)
  e^{-i\epsilon(k)t}.
\end{equation}
here $\epsilon(k)\simeq {k^2}/{2\mu}$ is the continuum state energy and $k$ is the corresponding asymptotic momentum.
The radial wavefunction can be written as
\begin{align}
  R_{k;\ell}(r)
  &=
  \frac{2k\,e^{\pi\mu\alpha/(2k)}
  \left|\Gamma\left(\ell+1+\frac{i\mu\alpha}{k}\right)\right|}
  {(2\ell+1)!}
  (2kr)^\ell e^{-ikr}
  \nonumber\\
  &\quad\times
  {}_1F_1\left(\ell+1+\frac{i\mu\alpha}{k};
  2\ell+2;2ikr\right).
\end{align}
For an orbit of period $T = 2 \pi/\Omega$, where $\Omega$ is the orbital frequency, 
the bound to continuum transition matrix element can be expanded as
\begin{equation}
  \langle k;\ell m|V(t)|b\rangle
  =
  \sum_{g\in{\mathbb Z}}
  \eta_{\ell m}^{(g)}(k)e^{-ig\Omega t}.
  \label{eq:ionfourier}
\end{equation}
here $\eta_{\ell m}^{(g)}(k)$ is the Fourier coefficient
\begin{equation}
  \eta_{\ell m}^{(g)}(k)
  =
  \frac{\Omega}{2\pi}
  \int_0^{2\pi/\Omega}\mathrm{d}t\,
  \langle k;\ell m|V(t)|b\rangle e^{ig\Omega t}.
  \label{eq:FourierCoefficient}
\end{equation}
Fermi's golden rule then gives the cloud depletion rate
\begin{equation}
  \Gamma_{\rm ion}
  =
  \sum_{\ell,m,g}
  \frac{\mu
  \left|\eta_{\ell m}^{(g)}(k^{(g)})\right|^2}
  {k^{(g)}}
  \Theta\!\left(k^{(g)2}\right) \, ,
  \label{eq:ionmdot}
\end{equation}
where $k^{(g)}$ is the outgoing momentum in the $g$-th harmonic channel. 
Energy conservation forces
\begin{equation}
  k^{(g)}
  =
  \sqrt{2\mu(\epsilon_b+g\Omega)} \, ,
  \label{eq:gchannelMomentum}
\end{equation}
where only harmonics with enough energy to reach the continuum contribute, i.e. $g\Omega > |\epsilon_{b}|$.
Using $N \simeq M_c/\mu$ in the non-relativistic limit,
the corresponding energy and angular momentum extracted from the orbit can be estimated as
\begin{align}
  P_{\rm ion}
  &=
  \frac{M_c}{\mu}
  \sum_{\ell,m,g} g\Omega\,
  \frac{\mu
  \left|\eta_{\ell m}^{(g)}(k^{(g)})\right|^2}
  {k^{(g)}}
  \Theta\!\left(k^{(g)2}\right),
  \label{eq:app-ion-power}\\
  \tau_{\rm ion}
  &=
  \frac{M_c}{\mu}
  \sum_{\ell,m,g} (m-m_b)\,
  \frac{\mu
  \left|\eta_{\ell m}^{(g)}(k^{(g)})\right|^2}
  {k^{(g)}}
  \Theta\!\left(k^{(g)2}\right).
  \label{eq:iontorque}
\end{align}

The same periodic perturbation $V(t,{\bf r})$ also couples the initial level to other bound state $|d\rangle$ with energy $\epsilon_d$.
For a weakly occupied final cloud state with decay rate $\gamma_d \equiv -2 \omega_i$, the cloud depletion rate from bound-bound transition can be written as
\begin{equation}
    \Gamma_{b \rightarrow d}^{(g)} 
    \simeq \frac{\gamma_d\left|
    \eta_{d b}^{(g)}\right|^2}{\left(\epsilon_d-\epsilon_b-g \Omega\right)^2+\left(\gamma_d / 2\right)^2},
\end{equation}
where $\eta^{(g)}_{db}$ is the corresponding bound-bound Fourier matrix element.
Such transitions are resonantly enhanced when an orbital harmonic closely matches a discrete level splitting $\epsilon_d-\epsilon_b=g\Omega$~\cite{Tomaselli:2023ysb,Tomaselli:2024bdd}. 
However, this restrictive frequency matching condition selects particular stellar orbits, 
rather than being generically satisfied. While off-resonant bound-bound transitions can also yield a meaningful contribution to the ionization rate, this contribution is more challenging to include, as the density of bound states near the continuum is large. Therefore, we focus exclusively on the bound-free transition rate, and emphasize once more that this is a lower limit to the rate of energy dissipation in the system.

\subsection{Depletion from Stellar Populations}

Let $\textbf{R}_o(t) = (R_o,\theta_o,\phi_o)$ denote the stellar orbital position. 
For a Keplerian orbit on the equatorial plane with semi-major axis $a$ and eccentricity $e$, 
the trajectory of the star is conveniently parametrized by the eccentric anomaly $E$. 
In particular, the instantaneous orbital radius is
\begin{equation}
 R_o(E) = a(1-e\cos E),
\label{eq:radialdistance}
\end{equation}
the remaining orbital variables satisfy
\begin{align}
\Omega t  &= E - e\sin E,\\
\tan\frac{\phi_o}{2}
&=\sqrt{\frac{1+e}{1-e}}\tan\frac{E}{2},\\
v_\star^2 & = 
G M \left( \frac{2}{R_o} - \frac{1}{a} \right)
\label{eq:true_anomaly}
\end{align}
where $\phi_o$ is the true anomaly measured from pericenter, $v_*$ denotes the stellar speed and $\Omega = \sqrt{GM/a^3}$ is the orbital frequency. 
Throughout this work, 
we calculate the depletion rates assuming for simplicity pro-grade equatorial orbits for each stellar population model. This makes the computation easier---though it could in principle be refined using the formulae for ionization on generic orbits obtained in Ref.~\cite{Xu:2026aic}. Note that pro-grade equatorial orbits are expected to yield near-maximal dissipation, while the actual maximum is generally higher by an order one factor, and is achieved for retro-grade orbits \cite{Tomaselli:2023ysb}.

Given the stellar number density distribution in semi-major axis $a$, eccentricity $e$, and stellar mass $M_\star$,
$d^3 N/ d a  d e d M_\star$,
one can compute the averaged cloud mass loss along an arbitrary eccentric orbit $\langle \Gamma_{\rm ion} \rangle$ by substituting the relations in Eq.~\eqref{eq:radialdistance} and \eqref{eq:true_anomaly} into Eq.~\eqref{eq:FourierCoefficient}. 
The total cloud depletion rate is then obtained by integrating over the stellar distribution,
\begin{equation}
\Gamma_{\rm tot}
= 
 \int d M_\star 
 \int_{r_t}^\infty da 
 \int_0^{1-\frac{r_t}{a}} de  \left<\Gamma_{\rm ion}\right> \frac{d^3 N}{da de dM_\star}
\label{eq:cloudmasslossEccen}
\end{equation}
Here $r_t \simeq R_\star
\left(M/M_\star \right)^{1/3} $ is the Newtonian tidal disruption radius for an ordinary star of mass $M_\star$ and radius $R_\star$. 
For compact remnants like BHs and neutron stars,
tidal disruption is negligible, 
and the orbital integration is instead restricted by the direct capture by the central BH.
In the Schwarzschild limit, 
the allowed orbits satisfy $a(1 - e) > 4GM$.

For sufficiently small BH masses, 
the tidal disruption radius lies well outside the event horizon. 
Since the characteristic size of a hydrogenic cloud scales as $r_c=GM/\alpha^2$,
increasing $\alpha$ makes the cloud more compact. 
Once the characteristic cloud radius lies inside the tidal disruption radius,
a stellar orbit would have to enter the tidal disruption region in order to perturb the cloud wave function. 
An ordinary star, 
however, cannot survive as an intact object in this region and therefore can no longer effectively probe the cloud.
In practice, we define the upper limit of the coupling $\alpha$ by requiring
$ r_t = 10 n r_c$, 
where the factor of $10$ is used to ensure that $r_t$ is placed in a region where the radial wave function is exponentially suppressed (and thus energy losses are negligible).  
This procedure gives
\begin{equation}
   \hspace{-2mm}
   \alpha_{\rm tidal}
   =
   0.731 \sqrt{n}
   \left( \frac{M}{4 \times 10^6 M_\odot } \right)^{1/3}
   \left(\frac{M_\star}{M_\odot}\right)^{1/6}
   \left(\frac{R_\star}{R_\odot}\right)^{-1/2}.
   \label{eq:alphamax}
\end{equation}

By analogy with the superradiant growth time $\tau_{\rm sr}$, we define the  corresponding cloud depletion timescale as
$\tau_{\rm loss} \equiv 1/\Gamma_{\rm tot}$.
The cloud mass then evolves, to linear order, as
\begin{equation}
    \frac{\dot M_c}{M_c} = \tau_{\rm sr}^{-1} - \tau_{\rm loss}^{-1}
\end{equation}
where we have neglected the contributions from gravitational wave emission and cloud self-interactions.
It follows from the equation above that, when $\tau_{\rm loss}/\tau_{\rm sr}\leq 1$, the stellar environment can prevent the superradiant growth of the cloud.

%
\section{Gravitational Dynamical Friction}
\label{sec:gravity}
In this section, we focus purely on gravitationally induced ionization/dynamical friction. 
This is an unavoidable effect arising from the local environment around any black hole, and thus it is crucial to understand how such effects alter the system, 
which parts of parameter space remain unaffected under pessimistic assumptions of the underlying stellar distributions, 
and what observations would be necessary in order to obtain a handle on the evolution of individual systems.

\subsection{Gravitational perturbation}

Let $q_\star \equiv M_\star/M$ denote the
stellar mass ratio. 
The orbiting stars induce a long range gravitational perturbation to the cloud. 
The corresponding Newtonian potential can be expanded in multipoles as
\begin{align}
  V_{\rm grav}(t,{\bf r})
  &=
  -\sum_{\ell_o m_o}
  \frac{4\pi\alpha q_\star}{2\ell_o + 1}
  Y_{\ell_o m_o}(\theta_o,\phi_o)
  Y^*_{\ell_o m_o}(\theta,\phi)
  \nonumber\\
  &\quad\times
  F_{\ell_o}(r;R_o),
\end{align}
where, for $\ell_o \ne 1$,
\begin{equation}
  F_{\ell_o}(r;R_o)
  =
  \frac{r^{\ell_o}}{R_o^{\ell_o + 1}}\Theta(R_o-r)
  +
  \frac{R_o^{\ell_o}}{r^{\ell_o + 1}}
  \Theta(r-R_o),
\end{equation}
and the dipole term is given by
\begin{equation}
  F_1(r;R_o)
  =
  \left(\frac{R_o}{r^2}-\frac{r}{R_o^2}\right)
  \Theta(r-R_o).
\end{equation}
The transition matrix element is
\begin{eqnarray}
  \langle k;\ell m|V_{\rm grav}(t)|b\rangle
  = \hspace{3cm} \\
  -\sum_{\ell_o m_o}
  \frac{4 \pi \alpha q_\star}{2\ell_o+1}
  Y_{\ell_o m_o}(\theta_o,\phi_o)
  I_r^{\rm grav} I_\Omega , \nonumber
  \label{eq:matrixeleGR}
\end{eqnarray}
with
\begin{align}
  I_r^{\rm grav}
  &=
  \int_0^\infty \mathrm{d}r\,r^2
  R^*_{k;\ell}(r)F_{\ell_o}(r;R_o)R_{n_b\ell_b}(r),
  \label{eq:ionirgrav}\\
  I_\Omega
  &=
  \int \mathrm{d}\Omega\,
  Y^*_{\ell m}Y^*_{\ell_o m_o}Y_{\ell_b m_b}.
\end{align}
The integral $I_\Omega$ can be expressed by using the Wigner-3j symbols and is nonzero only when the following conditions are satisfied:
\begin{align}
  -m-m_o+m_b&=0,\\
  \ell+\ell_o+\ell_b&\in 2{\mathbb Z},\\
  |\ell_b-\ell|&\le \ell_o\le \ell_b+\ell.
\end{align}
For an equatorial orbit, one has the additional constraint that
\begin{equation}
Y_{\ell_o m_o}\left(\frac{\pi}{2},0\right)\ne0,
\qquad
\ell_o+m_o\in2{\mathbb Z}.
\end{equation}
Substituting Eq.~\eqref{eq:matrixeleGR} into Eq.~\eqref{eq:FourierCoefficient} and \eqref{eq:cloudmasslossEccen}, 
one can obtain the gravitational ionization power $P^{\rm grav}_{\rm ion}$ and the total cloud depletion rate $\Gamma^{\rm grav}_{\rm tot}$.

The ionization process admits an equivalent local interpretation in terms of dynamical friction.
In the local limit, the bound cloud is approximated by a nearly uniform incident scalar wave, and the momentum carried away by the outgoing continuum states is interpreted as the drag exerted on the orbiting star.
The corresponding drag power for a star of mass $M_\star$ moving through the scalar cloud is~\cite{Hui:2016ltb}
\begin{equation}
P_{\rm DF}
=
\frac{4\pi G^2 M_\star^2 \rho}{v}\,
C_{\Lambda}(\xi,k r_{\Lambda}),
\label{eq:dynamicalFGR}
\end{equation}
where $v$ denotes the relative velocity between the star and the cloud. 
When the stellar speed $v_\star$ is much larger than the cloud rotation speed, $v \approx v_\star$.
The quantity $\rho$ denotes the local cloud density given in Eq.~\eqref{eq:threestatesrho}, 
while $C_{\Lambda}$ is an $\mathcal{O}(1)$ function of $\xi\equiv GM_\star\mu/v$ and $k r_{\Lambda}$, 
where $k=\mu v$ and the cutoff scale $r_{\Lambda}$ is defined as the smaller of the orbital radius $r$ and the characteristic cloud size.

Equation~\eqref{eq:dynamicalFGR} is a local approximation based on a
nearly uniform incident wave and rectilinear motion. It
therefore does not retain the discrete bound state spectrum or the harmonic thresholds in Eq.~\eqref{eq:gchannelMomentum}. Those features are  only resolved by the full ionization calculation. 
Nevertheless, the two descriptions capture the same transfer of orbital energy to scalar excitations and have the same parametric scaling. 
Explicit comparisons between the local dynamical friction calculation with the full ionization method are shown in Appendix~\ref{app:ComparisonIDF}. 
We therefore employ the full ionization treatment in our numerical calculations below. The 
dynamical friction description is only used as a simple estimate of gravitational depletion.

Using the dynamical friction formula in Eq.~\eqref{eq:dynamicalFGR}, 
we can identify which part of an eccentric orbit contributes most to the dissipation.
For gravitational dynamical friction, $P_{\rm DF} \propto \rho/v$ and 
for the three leading states considered in Eq.~\eqref{eq:threestatesrho}, 
the cloud density has the radial dependence $\rho_{n\ell m} \propto(r/r_c)^{2 \ell}\exp[-2r/(n r_c)]$.
At pericenter, $r_p=a(1-e)$, and apocenter, $r_a=a(1+e)$,
then the ratio of the contribution per unit eccentric anomaly is
\begin{equation}
\begin{aligned}
\frac{\left(P_{\mathrm{DF}} d t / d E\right)_p}{\left(P_{\mathrm{DF}} d t / d E\right)_a} &\propto  
\left(\frac{r_p}{r_a}\right)^{2\ell+2}
\exp  
\left[\frac{2 \left(r_a-r_p\right)}{nr_c}\right] \\
& = 
\left(\frac{1-e}{1+e}\right)^{2 \ell +2}
\exp  \left[\frac{4 a e}{n r_c}\right] .
\end{aligned}
\end{equation}
Here we have used $dt/dE = (1 - e \cos E )/\Omega \propto R_o$ and $v_p/v_a = r_a/r_p$, which follow directly from Eq.~\eqref{eq:radialdistance}.
For highly eccentric orbits, 
the power law factor varies much more slowly than the exponential cloud profile. 
Therefore, when $4 a e \gg  n r_c$, the energy loss is dominated by the pericenter passage.
In general, we evaluate energy loss for arbitrary orbits by computing the differential energy loss integrated along the orbital trajectory of any given star.
Given that the orbital period of the stars of interest is typically much shorter than the superradiant growth time,
we can freely work directly with the period averaged energy dissipation rate.

\subsection{Results}

We begin by comparing the net energy dissipation driven by the stellar population to the superradiant growth timescale of the three leading order superradiant modes (the $\left|211\right>$, $\left|322\right>$, and $\left|433\right>$ states). We perform this analysis for Sag A$^*$, and take an optimistic dimensionless spin of $\tilde{a} = 0.9$. 
We consider four stellar distributions, the observed S-star population~\cite{Tomaselli:2025zdo}, a single solar mass cusp profile, 
the strong segregation continuous stellar mass profile, and the multi-population model of Ref.~\cite{Generozov:2018niv}.

\begin{figure*}[!t]
    \centering
    \includegraphics[width=\textwidth]{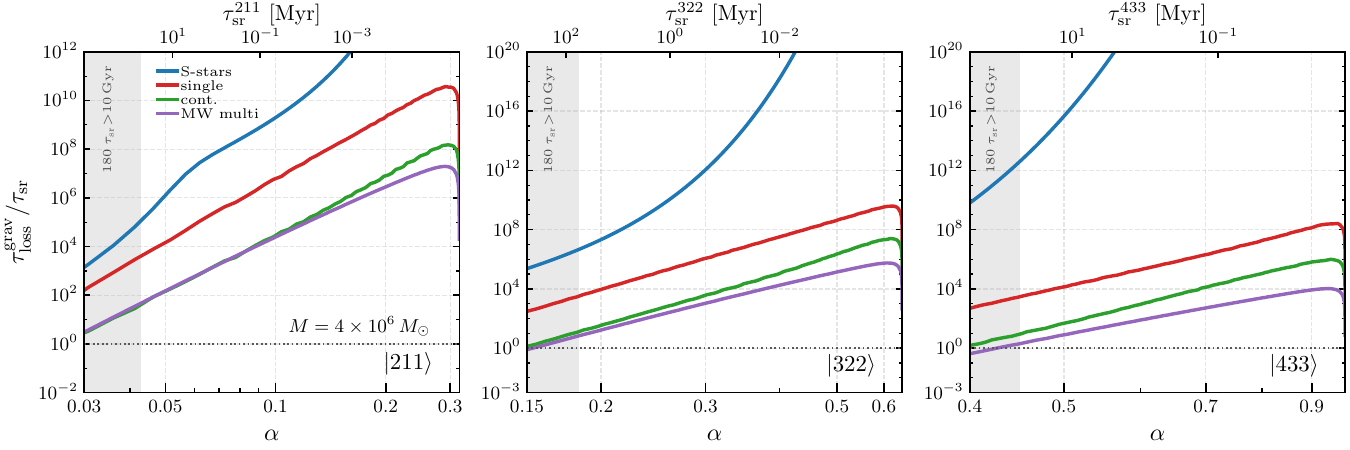}
    \caption{Ratio of the gravitational cloud mass depletion time to the superradiant growth time, $\tau^{\rm grav}_{\rm loss}/\tau_{\rm sr}$, for Sag A$^*$ with $M = 4 \times 10^6 M_\odot$.
    The three panels correspond to the $|211\rangle$, 
    $|322\rangle$ and $|433\rangle$ cloud states. 
    Blue, red, green and purple curves denote the S-stars population,
    the single-mass cusp, 
    and the strong segregated continuous mass profile, and the Milky way multi-population profile, respectively. 
    The upper axes give the corresponding superradiant growth time in Myr. The gray region marks $180 \, \tau_{\rm sr} = 10\,{\rm Gyr}$ and
    the horizontal dotted line marks $\tau^{\rm grav}_{\rm loss} = \tau_{\rm sr}$.
    The BH spin is taken to be $\tilde{a} = 0.9$.
    }
    \label{fig:gravity_only_ratio_mw}
\end{figure*}

Fig.~\ref{fig:gravity_only_ratio_mw} shows the ratio of the gravitational depletion time $\tau^{\rm grav}_{\rm loss}$ to the superradiant growth time $\tau_{\rm sr} = 1 / (2 \omega_i)$
for the Milky Way benchmark.
Each curve is truncated at both small and large $\alpha$,
the lower endpoint is set by the point at which $180 \times \tau_{\rm sr} \geq 10$ Gyr (since our comparison assumes a fully grown superradiant cloud), and the latter is set by
$\alpha_{\rm max} \equiv \min (\alpha_{\rm tidal}, \alpha_{\rm th})$. 
One finds that, as $\alpha$ increases, the cloud becomes more compact and the ratio generally increases.
One can also see from the figure that
the gravitational depletion can substantially slow down the cloud growth, 
particularly near the low-$\alpha$ boundary.
For the $|433\rangle$ state in the Milky Way multi-population model with $\alpha \approx 0.45$, 
$\tau_{\rm loss}^{\rm grav}/\tau_{\rm sr}\approx2$, implying that ionization removes approximately 50\% of the superradiant growth rate and thereby increases the net $e$-folding time.
For the $|322\rangle$ state with the same stellar profile, 
$\tau_{\rm loss}^{\rm grav}/\tau_{\rm sr}\approx8$, 
corresponding to an approximately 13\% reduction in the cloud growth rate.
The contribution from the observed S-stars alone is generally the weakest, giving $\tau^{\rm grav}_{\rm loss}\gg \tau_{sr}$.
This is because most S-star orbits lie well outside the typical radius of the superradiant cloud, $r_c$. 
The strong segregation continuous stellar mass profile and multi-population model give a shorter loss time, since stars and the heavy stellar remnants,
such as stellar-mass black holes, 
are more concentrated near the center and the drag force scales as $M_\star^2 n_\star$. 
The higher states are particularly sensitive to this effect because their cloud profiles extend to larger radii  and can overlap more efficiently with the segregated stellar population. 
Nevertheless, neither model competes with superradiant growth. Let us emphasize that a top-heavy stellar population, or a higher density of BHs, could easily enhance the rate of energy dissipation to the point where superradiant growth is prevented. Similarly, the inclusion of bound-bound transitions may already be sufficient within the existing models to have an important impact on these conclusions.

\begin{figure*}[!t]
    \centering
    \includegraphics[width=\textwidth]{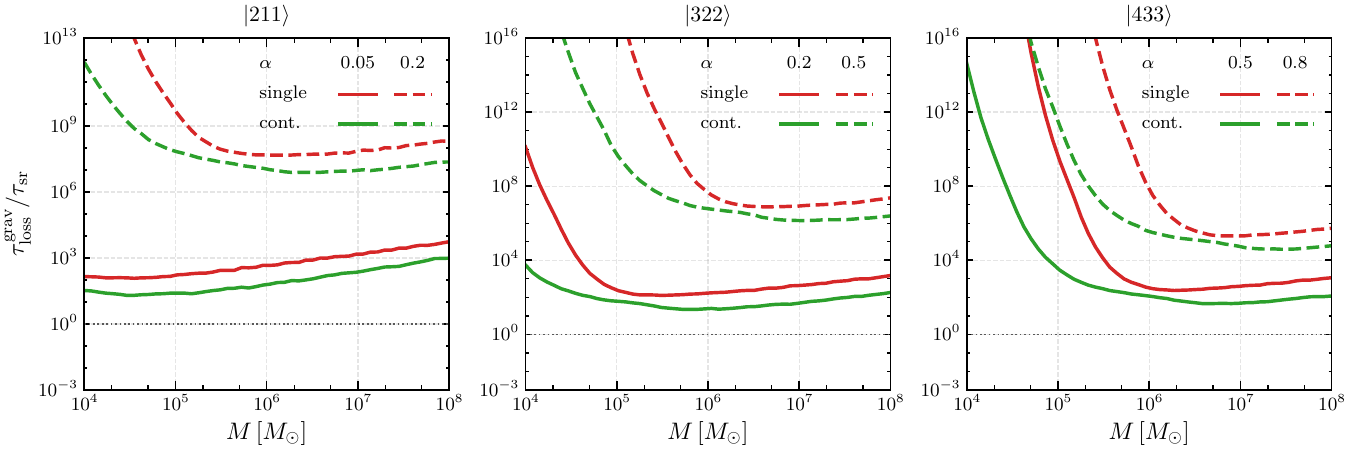}
    \caption{BH mass dependence of $\tau_{\rm loss }^{\rm grav } / \tau_{\rm sr }$ for the $|211\rangle$, $|322\rangle$ and $|433\rangle$ cloud states. 
    The red curves show the single solar-mass cusp, 
    and the green curves show the strongly segregated continuous stellar-mass profile.
    Solid and dashed curves correspond to $\alpha=0.05, 0.2, 0.5$ and $\alpha= 0.2,0.5,0.8$ for $|211\rangle,|322\rangle$ and $|433\rangle$ states,
    respectively. 
    The dotted horizontal line marks $\tau_{\rm loss }^{\rm grav}=\tau_{\rm sr}$. 
    The BH spin is taken to be $\tilde{a} = 0.9$.}
    \label{fig:gravity_only_ratio_mass_scan}
\end{figure*}

In Figure~\ref{fig:gravity_only_ratio_mass_scan}, we generalize this result to heavier and lighter BHs.
One can find that $\tau^{\rm grav}_{\rm loss}/\tau_{\rm sr}$ increases with higher BH mass $M$ for fixed $\alpha$. 
This is because larger $M$ makes the cloud radius grow as $r_c\sim G M/ \alpha^2$ such that it makes the cloud more dilute. 
In addition, from Eq.~\eqref{eq:dynamicalFGR}, the perturbation from an individual star is suppressed by factors of $q_\star^2= (M_\star / M)^2$. 
Although a heavier BH does enclose more stars inside the larger cloud, 
the total number of stars grows more slowly than $M^2$ and is not enough to compensate the $q_\star^2$ suppression, 
so gravitational depletion becomes less efficient relative to superradiant growth when increasing $M$.
As a simple numerical estimate, 
the single-mass cusp with $\gamma=1.4$ gives $N_{\rm cusp}(<r_c)\propto M^{1.57}$. 
Combining this with $q_\star^2\propto M^{-2}$ gives a collective perturbation strength proportional to $M^{-0.43}$, 
and hence $\tau_{\rm loss}^{\rm grav}/\tau_{\rm sr}\propto M^{0.43}$.
At the low BH mass end, ordinary stars are further affected by tidal disruption. 
The tidal cutoff removes them from the densest region of the cloud, 
where dynamical friction would otherwise be most efficient.

Our previous conclusion that the gravitational dissipation from bound-free transitions is unable to prevent superradiant growth remains unchanged over the BH mass range as shown in Figure~\ref{fig:gravity_only_ratio_mass_scan}. 
Since the depletion induced by an individual star scales approximately as $M_\star^2$, 
the total depletion rate is particularly sensitive to the abundance and spatial concentration of massive objects -- this can be directly seen by comparing the energy loss rates of the different stellar populations. Given that the rate of energy dissipation is in some cases comparable to the superradiant growth rate, future studies will benefit strongly from improved modeling of the local stellar populations. As discussed in Appendix~\ref{app:GWdepletion}, gravitational ionization {\emph{will}}  accelerate the decay of a saturated cloud, and is often far more efficient than GW emission -- this is significant since many observational probes pre-suppose the existence of a superradiant cloud of a particular mass.

%

\section{Dissipation from Quadratic Scalar-Matter Couplings}
\label{sec:couplings}

We consider a light scalar field $\phi$ with quadratic couplings to matter. 
Following the notation from~\cite{Day:2023mkb,Gan:2025icr}, the interaction Lagrangian is
\begin{equation}
  \mathcal{L}_{\rm int}=
  -\frac{\phi^2}{2 M^2_{\rm pl}} d_i \mathcal{O}^i_{\rm SM} .
\end{equation}
where $M_{\rm pl}$ is the reduced Planck mass, 
$d_i$ are dimensionless coupling coefficients, 
and $\mathcal{O}_{\rm SM}^i$ denote the relevant dimension-four Standard Model operators. 
In a stellar medium, these operators acquire nonzero expectation values,
$\langle \mathcal{O}^i_{\rm SM} \rangle = Q^i n({\bf r})$,
where $Q^i$ is the corresponding effective charge 
and $n({\bf r})$ is the number density of the medium. 
The scalar therefore acquires a position dependent in-medium mass,
\begin{equation}
\hspace{-1mm}
  m_{\rm med}^2({\bf r})
  =
  \mu^2 + \Delta m^2({\bf r}),
  \quad\;\ 
  \Delta m^2({\bf r})
  =
  \frac{d_i Q^i\, n({\bf r})}{M_{\rm pl}^2} \, .
  \label{eq:mediummassshift}
\end{equation}
Note that, in this work, 
we do not assume that $\phi$ constitutes the DM. 
Instead, the $\phi$ field is produced through BH superradiance.

\subsection{Scalar wave scattering off a localized target}

For a monochromatic wave $\phi({\bf r},t)=\mathrm{Re}[e^{-i\omega t}\psi({\bf r})]$, 
with vacuum dispersion relation $\omega^2 = k^2 + \mu^2$, 
the spatial wavefunction $\psi({\bf r})$ obeys
\begin{equation}
  \left[\nabla^2+k^2- \Delta m^2 ({\bf r})\right]\psi({\bf r})=0,
\end{equation}
Thus the sign of $d_i$ is physically important.  
For $d_i > 0$,
the stellar medium gives a positive mass shift and acts as a repulsive
potential barrier. If the barrier is strong enough that $\Delta m^2({\bf r})>k^2$, 
the wave is exponentially suppressed inside the medium and its penetration is
controlled by the skin depth
\begin{equation}
  L_{\rm skin}({\bf r})
  =
  \left[\Delta m^2({\bf r})-
  k^2\right]^{-1/2}
  \simeq \left[\Delta m^2({\bf r})\right]^{-1/2},
  \label{eq:skindepth}
\end{equation}
where the second equality assumes $\Delta m^2\gg k^2$.  
For $d_i < 0$, the medium instead produces an attractive well and can focus the incoming wave. 
Furthermore, if $m_{\rm med}^2 = \mu^2+\Delta m^2$ becomes negative in some part of the object, 
the assumed $\phi=0$ background becomes tachyonic and the object can source the scalar field~\cite{Hook:2017psm,Balkin:2020dsr,Balkin:2022qer}.

In this paper, we mainly focus on the case $d_i > 0$, for which the force is repulsive. For a spherically symmetric target of radius $R_*$,
the drag force is then obtained from the corresponding partial-wave scattering problem. 
Denoting $\delta_\ell$ the phase shifts induced by $V_{\rm eff}$, 
the time-averaged force on the target in a scalar background of energy density $\rho_\phi$ and velocity $v \simeq k/\mu$ is
\begin{equation}
  F_{\rm target}=\rho_\phi v^2 \sigma_T,
\end{equation}
with the momentum-transfer cross section
\begin{equation}
  \sigma_T
  =
  \frac{4\pi}{k^2}\sum_{\ell=0}^{\infty}(\ell+1)\sin^2(\delta_{\ell+1}
  -\delta_\ell),
\end{equation}

The behavior of the drag force is controlled by the hierarchy among $L_{\rm skin}$, 
the reduced de Broglie wavelength $\bar\lambda_{\rm dB}\equiv 1/k$, 
and the stellar radius $R_*$, 
leading to the simple limiting regimes discussed below.

\paragraph{Geometric / hard-sphere regime: $L_\mathrm{skin}\ll\bar\lambda_\mathrm{dB}\ll R_*$.} When the skin depth is much smaller than the de Broglie wavelength, the scalar field is effectively excluded from the stellar interior, 
so the target behaves as an opaque sphere. 
Since the wavelength remains much shorter than the target size, 
the scattering is well described by the classical hard-sphere limit,
and $\sigma_T$ reduces to the geometric area of the target. 
The resulting force is therefore
\begin{equation}
  \sigma_T = \pi R_*^2, \qquad
  F_\mathrm{target} = \rho_\phi\,v^2\,\pi R_*^2.
\end{equation}

\paragraph{$s$-wave regime: $L_\mathrm{skin}\ll R_*\ll\bar\lambda_\mathrm{dB}$.} If the potential remains strong but the de Broglie wavelength exceeds the stellar radius, the scattering
is dominated by the lowest partial wave. In this long-wavelength limit, the target still behaves as an
effectively hard object, but the cross section is now set by $s$-wave scattering rather than by classical
geometry. This leads to factor-of-four enhancement of the $\sigma_T$ relative
to the geometric result, and hence
\begin{equation}
  \sigma_T = 4\pi R_*^2, \qquad
  F_\mathrm{target} = 4\,\rho_\phi\,v^2\,\pi R_*^2.
  \label{eq:s-waveSigma}
\end{equation}

\paragraph{Born / weak-coupling regime: $R_*\ll L_\mathrm{skin}\ll\bar\lambda_\mathrm{dB}$.} In this regime the scalar field penetrates the target efficiently, and the interaction is sufficiently weak
that the Born approximation applies. Because the wavelength is much larger than the stellar radius, the
scattering is coherent over the entire target volume, 
so the amplitude is controlled by the volume integral of the effective potential.
The $\sigma_T$ is correspondingly suppressed relative to the
s-wave limit and is given by
\begin{equation}
  \begin{aligned}
    \sigma_T
    &=4\pi\left(\frac{\Delta m^2 R_*^3}{3}\right)^2,\\
    F_{\rm target}
    &=4\pi R_*^2\rho_\phi v^2
      \left(\frac{\Delta m^2 R_*^2}{3}\right)^2 .
  \end{aligned}
  \label{eq:bornregimeforce}
\end{equation}

\subsection{Ionization by quadratic coupling}

In this section,
we generalize the ionization formalism to the case of quadratic contact couplings.
For the parameter regime considered in this paper, 
the stellar radius $R_\star$ is much smaller than both the typical cloud radius $r_c$ and the outgoing scalar wavelength,  
$1/k^{(g)}$. 
The star is therefore unresolved by both the bound cloud and the outgoing scalar wave, 
and can be treated as a pointlike object.
The scattered wave is described by an amplitude $f(k,\theta)$, 
which determines the differential cross section through $d\sigma/d\Omega=|f(k,\theta)|^2$, where $\theta$ is the scattering angle.
For a spherically symmetric stellar potential, this amplitude can be decomposed into partial waves labeled by $\ell$.
Since $k R_\star \ll 1$,
higher partial waves are suppressed, and scattering is dominated by the angular independent $s$-wave.
Its amplitude is $f_0(k)=[k\cot\delta_0(k)-ik]^{-1}$, 
where $\delta_0(k)$ is the corresponding phase shift. 
At low energies, the scattering length $a_s$ encodes all short distance information at leading order, 
and the amplitude approaches $- a_s$ as $k \to 0$,
\begin{equation}
 a_{s}
 =-\lim_{k\to0}\frac{\tan\delta_0(k)}{k},
 \qquad
 \sigma_T(0)=4\pi a_{s}^2.
\end{equation}
The potential can therefore be replaced by the pseudopotential 
\begin{equation}
  V_{\rm ct}(t,{\bf r})
  =
  \frac{2 \pi a_s}{\mu}\,
  \delta^{(3)}({\bf r}-{\bf R}_o(t))
  \label{eq:pseudopotential} \, .
\end{equation}
For the object opaque to the scalar, 
which corresponds to the $s$-wave regime shown in Eq.~\eqref{eq:s-waveSigma}, the scattering length approaches $a_s \simeq R_\star$.
Conversely, if
the scalar can penetrate the object with $R_*\ll L_\mathrm{skin}$, one obtains the Born result, $a_s\simeq {\Delta m^2 R_\star^3}/{3}$.

The instantaneous bound-continuum matrix element is
\begin{align}
 \langle k;\ell m|V_{\rm ct}(t)|b\rangle
 ={}&\frac{2\pi a_{s}}{\mu}
 R^*_{k;\ell}\!\left[R_o(t)\right]
 R_{n_b\ell_b}\!\left[R_o(t)\right]
 \nonumber\\
 &\times
 Y^*_{\ell m}\!\left[\theta_o(t),\phi_o(t)\right]
 Y_{\ell_bm_b}\!\left[\theta_o(t),\phi_o(t)\right].
 \label{eq:quadraticcontactmatrix}
\end{align}
For a general co-rotating equatorial orbit, $\theta_o=\pi/2$, it is useful to keep the phase from the spherical harmonics
explicit,
\begin{align}
 Y^*_{\ell m}\!\left(\frac{\pi}{2},\phi_o\right)
 Y_{\ell_bm_b}\!\left(\frac{\pi}{2},\phi_o\right)
 ={}&Y^*_{\ell m}\!\left(\frac{\pi}{2},0\right)
 Y_{\ell_bm_b}\!\left(\frac{\pi}{2},0\right)
 \nonumber\\
 &\times e^{-i(m-m_b)\phi_o}.
 \label{eq:quadratic-angular-phase}
\end{align}
Substituting Eq.~\eqref{eq:quadraticcontactmatrix} into
Eq.~\eqref{eq:FourierCoefficient} gives the Fourier coefficient for an eccentric equatorial orbit 
\begin{align}
 \eta_{\ell m,{\rm ct}}^{(g)}(k)
 ={}&\frac{\Omega a_{s}}{\mu}
 Y^*_{\ell m}\!\left(\frac{\pi}{2},0\right)
 Y_{\ell_bm_b}\!\left(\frac{\pi}{2},0\right)
 \nonumber\\
 &\times\int_0^{2\pi/\Omega}dt\,
 R^*_{k;\ell}\!\left[R_o(t)\right]
 R_{n_b\ell_b}\!\left[R_o(t)\right]
 \nonumber\\
 &\times
 \exp\!\left[ig\Omega t-i(m-m_b)\phi_o(t)\right].
 \label{eq:quadraticcontactfourier}
\end{align}
The contact operator does not generate a Gaunt integral and hence has no
triangle selection rule. 
The nonzero equatorial matrix elements require
\begin{equation}
 |m|\leq\ell,
 \qquad \ell+m\in2{\mathbb Z},
 \qquad \ell_b+m_b\in2{\mathbb Z}.
 \label{eq:quadraticcontactselection}
\end{equation}
For a circular co-rotating orbit with $R_o(t)=R_0$ and
$\phi_o(t)=\Omega t$,  Eq.~\eqref{eq:quadraticcontactfourier} further reduces to
\begin{align}
 \eta_{\ell m,{\rm ct}}^{(g)}
 \!\left(k^{(g)}\right)
 ={}&\frac{2\pi a_{s}}{\mu}
 R^*_{k^{(g)};\ell}(R_0)R_{n_b\ell_b}(R_0)
 \nonumber\\
 &\times
 Y^*_{\ell m}\!\left(\frac{\pi}{2},0\right)
 Y_{\ell_bm_b}\!\left(\frac{\pi}{2},0\right)
 \delta_{g,m-m_b}.
 \label{eq:quadraticcontactcircular}
\end{align}
where the delta function imposes an additional frequency matching condition $g=m-m_b$.  

The corresponding ionization power can be obtained directly from
Eq.~\eqref{eq:app-ion-power}, and the total cloud depletion rate is
obtained by inserting the resulting single orbit rate into
Eq.~\eqref{eq:cloudmasslossEccen}.

\subsection{Quadratic coupling to nucleons}

We begin by considering the quadratic coupling to nucleons, 
\begin{equation}
  \mathcal{L}_\mathrm{int} = -\frac{d_N m_N}{2 M^2_{\rm pl}}\,\phi^2\,\bar{N}N.
\end{equation}
Inside a macroscopic object one may replace the nucleon bilinear by the local
number density, $\left< \bar N N \right> \simeq n_N({\bf r}) \simeq \rho({\bf r})/m_N$, 
where $\rho({\bf r})$ is the density of the nucleon matter.
Using Eq.~\eqref{eq:mediummassshift},
the nucleon induced mass shift is $\Delta m_N^2 = d_N m_ N n_N({\bf r})/M^2_{\rm pl} \simeq d_N \rho(r)/M_{\rm pl}^2$.
For the repulsive case with $d_N>0$,  and when $\Delta m_N^2\gg k^2$, the skin depth is roughly given by
\begin{equation}
  L_{\rm skin}^{(N)}({\bf r})
  \simeq
  \frac{M_{\rm pl}}{\sqrt{d_N m_N n_N({\bf r})}} \, ,
  \label{eq:nucleonskindepth}
\end{equation}
which numerically corresponds to characteristic values of
\begin{equation}
  L_{\rm skin}^{(N)}
  \simeq
  2.3 \times 10^{-1}\,{\rm km}
  \left(\frac{10^{18}}{d_N}\right)^{1/2}
  \left(\frac{1\,{\rm g\,cm}^{-3}}{\rho}\right)^{1/2} .
  \label{eq:nucleonskindepthnumeric}
\end{equation}
The most stringent constraint on repulsive quadratic scalar-nucleon couplings, and which does not make any assumption on the nature of dark matter, comes from SN1987A energy loss bounds on the process $N N \rightarrow N N \phi \phi$, which implies $d_N \lesssim 5.6 \times 10^{28}$~\cite{Sibiryakov:2020eir}. For the benchmark coupling $d_N=10^{18}$, 
the corresponding skin depths evaluated at various relevant densities are given by: 
\begin{eqnarray}
L_{\rm skin}^{(N)}&\simeq&2.0\times10^{-1}\,{\rm km} \hspace{.5cm} \rho\simeq1.4\,{\rm g\,cm}^{-3} \\
L_{\rm skin}^{(N)}&\simeq& 1.9\times10^{-2}\,{\rm km} \hspace{.5cm} \rho\simeq150\,{\rm g\,cm}^{-3} \\
L_{\rm skin}^{(N)} &\simeq& 2.3\times10^{-4}\,{\rm km} \hspace{.5cm} \rho\simeq10^6\,{\rm g\,cm}^{-3} \\ 
L_{\rm skin}^{(N)} &\simeq& 1.3 \times 10^{-8}\,{\rm km} \hspace{.5cm} \rho\simeq3\times10^{14}\,{\rm g\,cm}^{-3} 
\end{eqnarray}
where the characteristic densities correspond to mean solar density, typical solar core density, characteristic white dwarf density, 
and neutron star core density.
These microscopic penetration
lengths imply that scattering off typical stellar objects is in the $s$-wave regime for a $d_N$ even well below the current SN1987A bound.
Solving for the coupling at which $L^{(N)}_{\rm skin} \sim R_\star$, one finds
\begin{equation}
    d_N ( R_\star)
    \equiv \frac{M_{\rm pl}^2}
    {\rho R_{\star}^2} 
    \simeq 1.1 \times 10^5\left(\frac{1\, \mathrm{g} \,\mathrm{cm}^{-3}}{\rho}\right)\left(\frac{R_{\odot}}{R_{\star}}\right)^2
\end{equation}
Thus $d_N \ll d_N (R_\star)$ gives penetrative Born scattering, whereas $d_N \gg d_N (R_\star)$ gives an opaque target.
For a more accurate calculation of the scattering cross section, 
we use stellar profiles generated with \texttt{MESA} and compute the phase shifts by solving the Schrödinger equation. 
The stellar profiles and transport cross sections used in this calculation are summarized in Fig.~\ref{fig:stellar_density_profiles} and discussed in Appendix~\ref{app:StellarProfile}.

\begin{figure*}[!t]
    \centering
    \includegraphics[width=\textwidth]{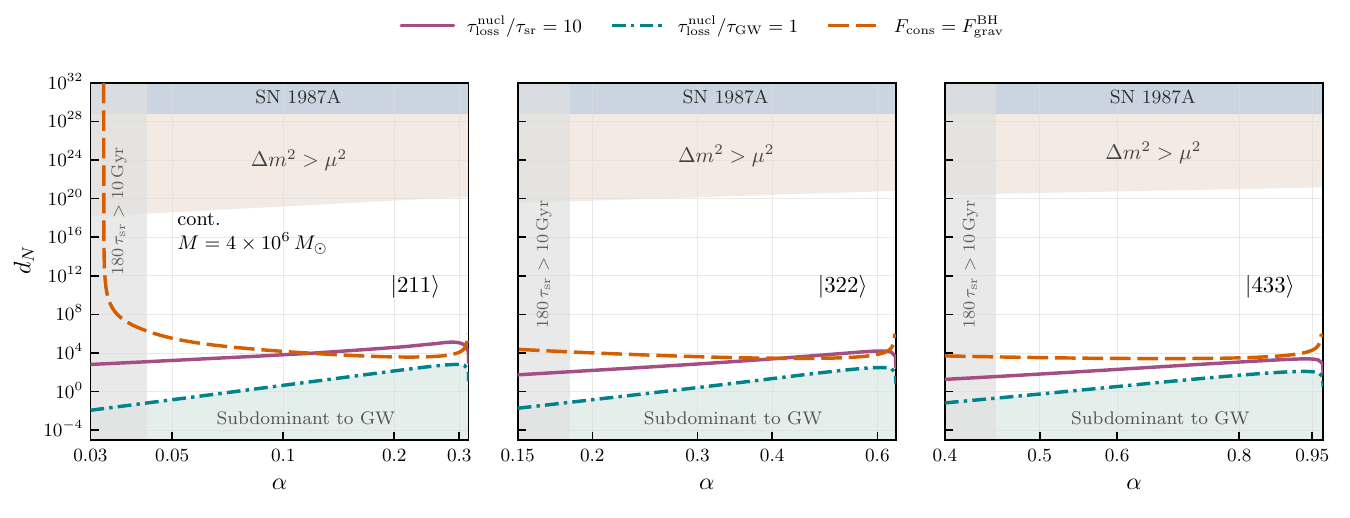}
    \caption{Parameter space of interest for the repulsive quadratic scalar–nucleon coupling, 
    using the strongly segregated continuous stellar population
    with $M=4\times10^6\,M_\odot$.
    The panels show results for the $|211\rangle$, $|322\rangle$, and $|433\rangle$ states, respectively.
    The purple curves indicate $\tau^{\rm nucl}_{\rm loss}/\tau_{\rm sr}=10$. 
    The green curves represent $\tau^{\rm nucl}_{\rm loss}/\tau_{\rm GW}=1$, 
    while the orange dashed curve marks where the smallest coupling for which the conservative force exerted by the cloud on a solar-mass star matches that of gravity.
    Both the green and orange curves are evaluated at the cloud's saturation mass (for smaller cloud masses, the orange curve shifts up, while the green curve shifts down).
    In the green shaded region, stellar depletion is subdominant to GW emission.
    The light brown region indicates $\Delta m_N^2 > \mu^2$ for diffuse matter with $\rho_N=10^{-18}\,{\rm g\,cm^{-3}}$ around Sgr A$^\star$. 
    The blue gray shading shows the SN1987A bound, $d_N=5.6\times10^{28}$ and the gray shaded regions satisfy $180\,\tau_{\rm sr}>10\,{\rm Gyr}$.}
    \label{fig:dN_alpha_mw}
\end{figure*}

Using Eq.~\eqref{eq:cloudmasslossEccen}, and including the cloud depletion rate induced by the quadratic nucleon coupling, 
one can calculate the cloud depletion time
$\tau^{\rm nucl}_{\rm loss}$.

Figure~\ref{fig:dN_alpha_mw} provides an overview of the parameter space of interest for the repulsive coupling $d_N>0$. 
For the strongly segregated continuous stellar mass distribution adopted for the Milky Way, 
the purple curve marks $\tau^{\rm nucl}_{\rm loss}/\tau_{\rm sr}=10$, 
where the stellar depletion rate reaches around 10\% of the superradiant growth rate. 
Below this threshold, depletion has a negligible effect on cloud growth. 
The green dash-dotted
curve marks $\tau^{\rm nucl}_{\rm loss}/\tau_{\rm GW}=1$, 
where stellar depletion matches the gravitational wave loss rate of a cloud at its saturation mass.
The light brown shaded region corresponds to an order-one or larger correction to the scalar mass squared, $\Delta m^2>\mu^2$, 
induced by diffuse matter around the BH, for which we adopt a benchmark density of $\rho_N=10^{-18}\,{\rm g\,cm^{-3}}$, which is consistent with models of the inner accretion flow around Sgr A$^\star$~\cite{Genzel:2010zy,Cho:2021wbk}.
This boundary serves as an illustrative reference for substantial medium effects, 
which can shift the cloud's energy levels and  potentially change the stability of an initially growing mode.
The blue gray shaded region shows the SN1987A bound.
Our main region of interest lies above the green curve and below the light brown shaded region. 
Here, stellar scattering can shorten the cloud lifetime and, 
above the purple curve, 
appreciably suppress cloud growth, 
while the matter induced scalar mass shift remains small enough to preserve the standard qualitative picture of superradiant cloud evolution.

The orange dashed curve provides a benchmark for the effect of the cloud’s conservative force $F_{\rm cons}$ on stellar orbits.
Following Appendix~\ref{app:ConservativeF}, we evaluate the $F_{\rm cons} = - N_b(t) \, \nabla \left<b | V | b \right>$ using the contact potential in Eq.~\eqref{eq:pseudopotential}, and compare its magnitude with the gravitational force from the BH, $F_{\rm grav}^{\rm BH}= G M M_\star/r^2$.
When the two forces become comparable,
the cloud can induce appreciable deviations from Keplerian stellar motion. We evaluate this effect for a fully saturated cloud, and at the characteristic cloud radius, in order to highlight the minimal coupling for which this can be relevant -- for smaller cloud masses, this curve shifts upward.
For small values of $\alpha$, the reduced saturation mass and greater spatial extent of the cloud suppress the conservative force such that this curve always lies above the purple boundary (corresponding to the coupling where stellar interactions modify superradiant growth). At higher values of $\alpha$ this is not always the case; however, $F_{\rm cons}\propto M_c$, 
so reducing the cloud mass weakens the orbital perturbation and shifts the orange curve upward. 
Since the fractional energy dissipation rate is independent of the cloud mass, the purple boundary remains unchanged to such a shift in the cloud mass. 

\begin{figure*}[!t]
    \centering
    \includegraphics[width=\textwidth]{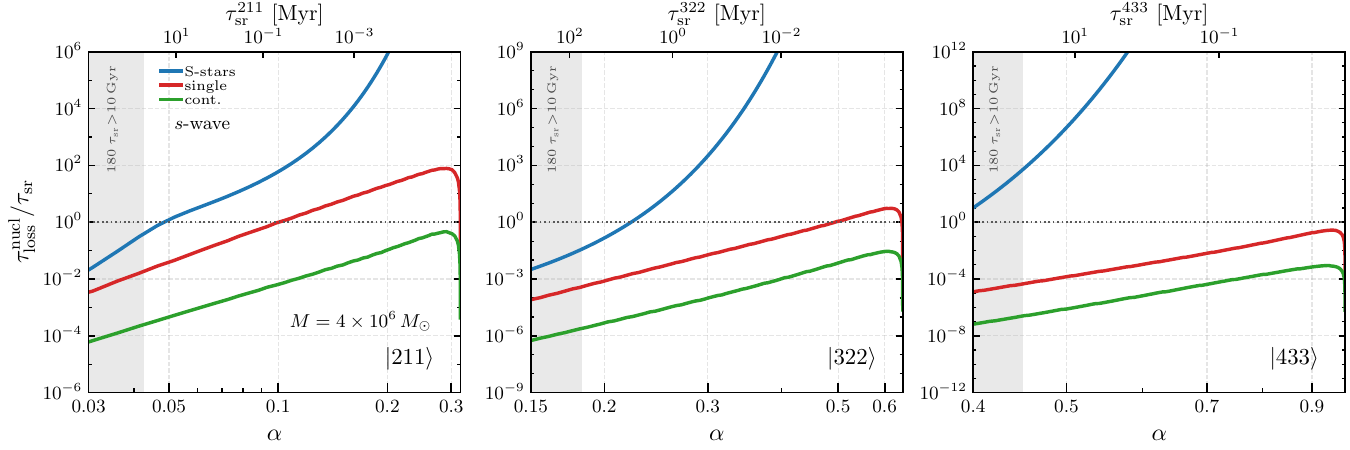}
    \caption{
    The stellar depletion for quadratic scalar nuclear coupling in the Milky Way benchmark. We show
    $\tau^{\rm nucl}_{\rm loss}/\tau_{\rm sr}$ for s-wave scattering
    as a function of $\alpha$ for
    $M=4\times10^6 M_\odot$ and $\tilde{a}=0.9$. The three panels correspond to the
    $|211\rangle$, $|322\rangle$, and $|433\rangle$ cloud profiles. 
    The red curves denote the single solar-mass cusp, 
    blue curves denote the observed
    S-star population and the green curves denote the strong-segregated continuous stellar-mass profile. 
    The gray shaded region
    marks $180\,\tau_{\rm sr}>10\,{\rm Gyr}$.
    } 
    \label{fig:quartic_ratio_mw_swave}
\end{figure*}

Figure~\ref{fig:quartic_ratio_mw_swave} shows the result for the Milky Way benchmark, adopting a value of $d_N$ near the maximally allowed value.  
For both the single solar-mass cusp and the strong segregated continuous stellar-mass distribution,
the depletion rate can give $\tau^{\rm nucl}_{\rm loss}/\tau_{\rm sr}<1$ over a broad range of $\alpha$. 
This means that in this regime the stellar population depletes the cloud faster than superradiance can replenish it, 
and can therefore strongly affect the growth of the cloud.  
The strong segregation continuous mass function gives the shortest loss time  -- this is because it contains a higher number density of stellar objects, and more massive stars, both of  
which increase the effective cross section near the cloud. 
The S-star population is much less effective, 
since only a small number of observed stars pass through the radii within the scale of $r_c$.

\begin{figure*}[!t]
    \centering
    \includegraphics[width=\textwidth]{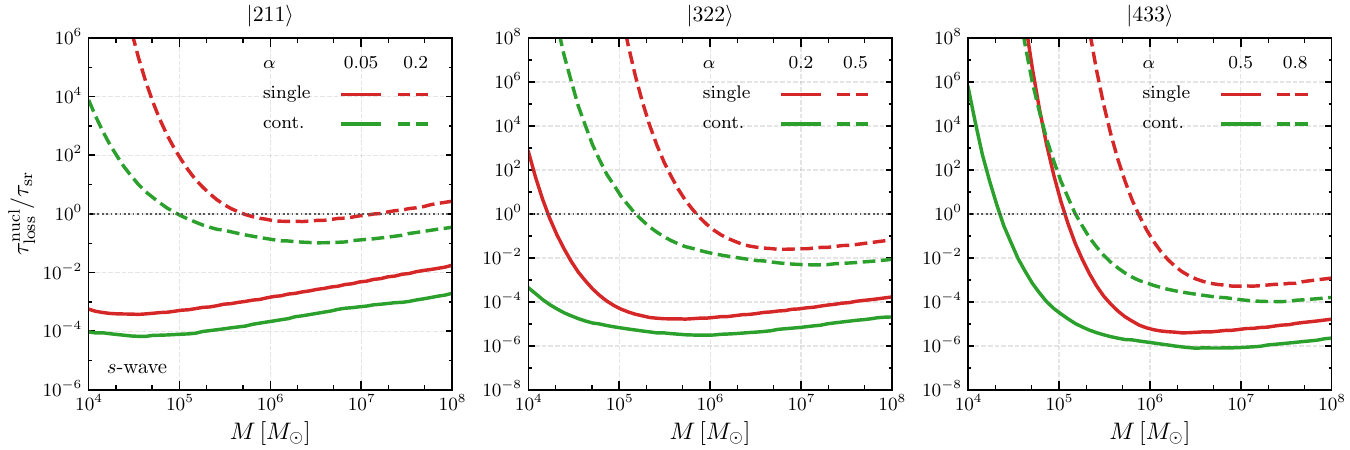}
    \caption{
    BH mass dependence of $\tau^{\rm nucl}_{\rm loss}/\tau_{\rm sr}$ for quadratic scalar nuclear coupling in the s-wave limit.
    The three panels correspond
    to the $|211\rangle$, $|322\rangle$, and $|433\rangle$ clouds, respectively.
    Red curves show the single solar-mass cusp, while green
    curves show the strongly segregated continuous stellar mass profile. 
    In the three panels, the solid curves use $\alpha = 0.05, 0.2$, and $0.5$, respectively, while the dashed curves use $\alpha= 0.2, 0.5$, and $0.8$.
    The horizontal dashed line marks the point at which the depletion
    time equals the superradiant growth time.  
    The BH spin is $\tilde a = 0.9$.
    } 
    \label{fig:quartic_ratio_mass_scan_swave}
\end{figure*}

In Fig.~\ref{fig:quartic_ratio_mass_scan_swave}, we generalize the calculation to heavier and lighter black holes.
For each cloud state, 
we show two representative values of $\alpha$ and compare the single solar-mass cusp with the strong segregated continuous stellar mass profile. 
Over most of the mass range, 
decreasing the BH mass makes the scalar cloud more susceptible to stellar drag depletion. 
This is mainly because the characteristic cloud radius, 
$r_c=GM/\alpha^2$, moves inward as $M$ decreases,
while the stellar density increases at smaller radii. 
Both effects enhance the orbit averaged energy loss rate, 
thereby reducing $\tau^{\rm nucl}_{\rm loss}/\tau_{\rm sr}$.

For small BH masses, $M < 10^5 M_\odot$, 
particularly for the larger values of $\alpha$, 
however, the trend reverses and $\tau^{\rm nucl}_{\rm loss}/\tau_{\rm sr}$ rises sharply. 
This happens because the cloud radius becomes so small that the region of largest scalar density lies inside the tidal disruption radius.
Stars can therefore no longer pass through the peak of the cloud, 
leading to the sudden increase in $\tau^{\rm nucl}_{\rm loss}/\tau_{\rm sr}$.


Let us briefly note that Ref.~\cite{Yu:2026ebx} also studied the effect of a quadratic scalar nucleon coupling in the context of superradiance around Sag A$^*$; this work focused on how such an interaction would affect the orbit of S2, but did not discuss whether stellar populations more generally alter the growth or decay of the superradiant cloud.

\subsection{Quadratic coupling to electromagnetism}

We next consider the quadratic coupling to electromagnetism,
\begin{equation}
  \mathcal{L}_\gamma
  =
  -\frac{d_\gamma}{4M_{\rm pl}^2}\phi^2F_{\mu\nu}F^{\mu\nu} .
  \label{eq:photon_quadratic_coupling}
\end{equation}
where $F_{\mu\nu}F^{\mu\nu}=2(B^2-E^2)$.
This interaction shifts the electromagnetic fine-structure constant by
$\delta\ln\alpha_{\rm em}=-d_\gamma\phi^2/M_{\rm pl}^2$,
thereby changing the electromagnetic contributions to the masses of ordinary matter.
For an atom with mass $m_i$,
this change is characterized by the effective charge $Q_{\gamma,i}\equiv\partial\ln m_i/\partial\ln\alpha_{\rm em}$.
For a nucleus with charge $Z$ and mass
number $A$, 
nucleon electromagnetic self-energies and nuclear Coulomb energy give~\cite{Damour:2010rp,Gan:2025nlu}
\begin{equation}
 Q_\gamma^{(A,Z)}
 \simeq10^{-4}
 \left[-1.4 + 8.2\frac{Z}{A} 
 + 7.7\frac{Z(Z-1)}{A^{4/3}}
 \right],
 \label{eq:photon_nuclear_charge}
\end{equation}
For a mixture of matter density $\rho$ and mass fractions $X_i$, 
the effective charge averaged over the mixture is $Q_{\gamma,\rm eff} = \sum_iX_iQ_{\gamma,i}$,
with $\sum_i X_i=1$.

At fixed particle number densities, 
the $\phi$-dependent masses change the matter energy density, thereby contributing to the scalar effective potential.
Because this contribution is quadratic in $\phi$, 
it shifts the scalar mass. 
Including the contribution from macroscopic electromagnetic fields, we obtain
\begin{equation}
 \Delta m_\gamma^2({\bf r})
 \simeq\frac{d_\gamma}{M_{\rm pl}^2}
 \left[B^2-E^2-2Q_{\gamma,\rm eff}\rho\right].
 \label{eq:photonmassshift}
\end{equation}
A repulsive potential gives the positive mass shift.
For an incident scalar of wave number $k$, a region with $\Delta m_\gamma^2>k^2$ has penetration length
\begin{equation}
 L_{\rm skin}^{(\gamma)}
 =\left(\Delta m_\gamma^2-k^2\right)^{-1/2}
 \simeq\left(\Delta m_\gamma^2\right)^{-1/2},
 \label{eq:photonskindepth}
\end{equation}
where the last approximation requires $\Delta m_\gamma^2\gg k^2$.

\paragraph{Stars.}
\label{par:stars}

For ordinary stars, where the matter contribution dominates, $d_\gamma > 0$ gives an attractive potential. 
Such interactions can produce resonances or stellar instabilities, 
leading to strong astrophysical constraints~\cite{Hook:2017psm,Balkin:2022qer,Gomez-Banon:2024oux}.
We therefore focus on $d_\gamma<0$,
which gives a repulsive potential and adopt a solar-like composition with $X_{\rm H}=0.70$, $X_{\rm He}=0.28$, 
and a metal mass fraction of $0.02$,
mainly oxygen, carbon, and iron. 
For such a composition, Eq.~\eqref{eq:photon_nuclear_charge}  gives $Q_{\gamma,\rm eff}\simeq6.5\times10^{-4}$, 
with hydrogen and helium contributing $6.2\times10^{-4}$, and the metals supplying a few percent correction. 
These nuclear contributions can also be applied to the ionized plasma of the stars, since the atomic binding and plasma interaction energies give only sub-leading corrections.
For the mean solar density, 
$\rho\simeq1.4\,{\rm g\,cm^{-3}}$, 
and a benchmark internal stellar magnetic field, $B_{\rm inv}\equiv\sqrt{B^2-E^2}=10^6\,{\rm G}$~\cite{OHare:2020wum}, 
the magnetic contribution to the shift in mass is about $5\times10^{-8}$ times smaller than the contribution from matter -- we therefore neglect it in the rest of the computation.
Since the remaining contribution is proportional to the matter density $\rho$, one can directly use the results of the proceeding section on quadratic scalar-nucleon couplings by using  $\Delta m_N^2=d_N\rho/M_{\rm pl}^2$, which yields a relation
\begin{equation}
 d_N^{\rm eff}({\bf r})=-2Q_{\gamma,\rm eff}({\bf r})d_\gamma.
 \label{eq:photon_nucleon_conversion}
\end{equation}
For the adopted composition, $d_N^{\rm eff}\simeq-1.3\times10^{-3}d_\gamma$.
Neglecting macroscopic electromagnetic fields, 
this substitution yields the same scattering potential for a fixed stellar density profile. 
Thus, for ordinary stars, results for 
the quadratic coupling to electromagnetism map directly onto those for the quadratic coupling to nucleons.

For $\Delta m_\gamma^2\gg k^2$,
the characteristic penetration length is
\begin{equation}
 \begin{split}
 L_{\rm skin}^{(\gamma)}
 &\simeq\frac{M_{\rm pl}}{\sqrt{2|d_\gamma|Q_{\gamma,\rm eff}\rho}}\\
 &\simeq0.78R_\odot
 \left(\frac{10^8}{|d_\gamma|}\right)^{1/2}
 \left(\frac{6.5\times10^{-4}}{Q_{\gamma,\rm eff}}\right)^{1/2}\\
 &\hspace{8mm}\times
 \left(\frac{1.4\,{\rm g\,cm^{-3}}}{\rho}\right)^{1/2}.
 \end{split}
 \label{eq:photonpenetration}
\end{equation}
At this mean density, the coupling constants $d_\gamma=-10^{12},-10^{18},-10^{24}$ correspond to skin depths of
$L_{\rm skin}^{(\gamma)}\simeq5.4\times10^3\,{\rm km}$,
$5.4\,{\rm km}$, and $5.4\,{\rm m}$, respectively. The coupling at which the penetration length equals the stellar radius $R_\star$ is given by
\begin{equation}
 \begin{split}
 d_\gamma(R_\star)
 \simeq{}& - 6.1 \times 10^7
 \left(\frac{6.5\times10^{-4}}{Q_{\gamma,\rm eff}} \right)\\
 &\times
 \left(\frac{1.4\,{\rm g\,cm^{-3}}}{\bar\rho}\right)
 \left(\frac{R_\odot}{R_\star}\right)^2,
 \end{split}
 \label{eq:dgamma-regime-boundaries}
\end{equation}
where $\bar\rho$ is the mean stellar density. This value corresponds to the transition between weak scattering and the regime where the scalar field is strongly expelled from the star.

\paragraph{Neutron stars.}
\label{par:neutron-stars}

In the case of neutron stars, the strong magnetic fields can overcome this suppression and dominate over the matter contribution.  Here, we consider the external magnetosphere, 
neglecting the matter contribution and assuming $B^2-E^2>0$ over the radii relevant for scattering. 
The mass shift is then
$\Delta m_\gamma^2\simeq d_\gamma(B^2-E^2)/M_{\rm pl}^2$.
The magnetic potential is repulsive for $d_\gamma>0$, 
opposite to the ordinary star branch.

The light cylinder of a neutron star with angular frequency $\Omega$ is given by 
\begin{equation}
  R_{\rm LC}
  =\frac{1}{\Omega}
  =\frac{P}{2\pi}
  =4.77\times10^3\,{\rm km}
  \left(\frac{P}{0.1\,{\rm s}}\right),
  \label{eq:light-cylinder}
\end{equation}
where $P$ is the spin period.  Inside the light cylinder, we will approximate the  magnetic field as a co-rotating dipole -- while higher order moments can exist, the spin-down power provides an estimate of the magnetic field on large distances, where the dipolar component should dominate, implying the dipolar treatment is conservative. For simplicity, we take an aligned rotator, with the magnetic field structure thus given by
\begin{equation}
  \begin{aligned}
    B_r
    &=B_p\left(\frac{R_\star}{r}\right)^3\cos\theta,\\
    B_\theta
    &=\frac{B_p}{2}\left(\frac{R_\star}{r}\right)^3\sin\theta,\\
    B_\varphi&=0,
  \end{aligned}
  \label{eq:NS-dipole-field}
\end{equation}
with magnitude
\begin{equation}
  B^2
  =
  \frac{B_p^2}{4}
  \left(\frac{R_\star}{r}\right)^6
  \left(1+3\cos^2\theta\right) .
  \label{eq:NS-dipole-magnitude}
\end{equation}
where $B_p$ is the surface field value at the pole.  The co-rotation vacuum electric field in the
closed magnetosphere satisfies $E/B\sim\Omega r\sin\theta$, and is in general screened in the inner magnetosphere -- as such, we take  $B^2-E^2\simeq B^2$.

Outside the light cylinder, the magnetic field lines open, and the structure of the field tends toward a configuration that resembles the split-monopole ~\cite{Timokhin:2006,Timokhin:2007xj,Tchekhovskoy:2015cda}.
We parametrize the radial component by the open magnetic flux,
\begin{equation}
  B^{\rm wind}_r(r)
  =
  \frac{\Phi_{\rm open}}{2\pi r^2}
  \label{eq:split_monopole_wind}
\end{equation}
Here $\Phi_{\rm open}$ is the open magnetic flux carried by one hemisphere of the wind. Fiducially, we take the open flux to be
\begin{equation}
  \Phi_{\rm open}
  =
  \kappa_\Phi
  \frac{2 \pi \mu_m}{R_{\rm LC}},
  \qquad
  \mu_m=\frac{1}{2} B_p R_\star^3 .
  \label{eq:open_flux_wind}
\end{equation}
Here $\mu_m$ is the stellar magnetic dipole moment, and $2 \pi \mu_m/R_{\rm LC}$ is the dipolar magnetic flux associated with a single
hemisphere up to the flux surface whose equatorial crossing is at the light
cylinder.  
The dimensionless parameter $\kappa_\Phi$ accounts for the uncertainty in how much of the dipolar polar cap flux is opened by the magnetosphere. 
In the numerical estimates below we use the fiducial value, $\kappa_\Phi=1.23$~\cite{Timokhin:2006,Timokhin:2007xj}.
The toroidal and electric components scale as
\begin{equation}
  \begin{aligned}
    B^{\rm wind}_\varphi
    &\simeq 
    -\frac{\Omega r\sin\theta}{c}
    B^{\rm wind}_r,\\
    E^{\rm wind}_\theta&\simeq -B^{\rm wind}_\varphi,
    \qquad E^{\rm wind}_r=E^{\rm wind}_\varphi 
    = B^{\rm wind}_\theta =0 .
  \end{aligned}
  \label{eq:NS-wind-fields}
\end{equation}
Thus in the far wind zone, we approximate the field structure as
\begin{equation}
  B^2-E^2
  \simeq (B^{\rm wind}_r)^2
  =
  \left(B_r^{\rm wind} 
  (R_{\rm LC})\right)^2
  \left(\frac{R_{\rm LC}}{r}\right)^4 .
  \label{eq:wind_invariant}
\end{equation}
%


Normal pulsars typically have surface fields of order
$10^{11}$--$10^{12}\,{\rm G}$, whereas magnetars have fields of order
$10^{14}$--$10^{15}\,{\rm G}$~\cite{Bransgrove:2018ns,Kaspi:2017fwg}.  At the
neutron star surface, the electromagnetic penetration length is
\begin{equation}
  L_{\rm skin}^{(\gamma)}(R_\star)
  \simeq
  78\,\mu{\rm m}
  \left(\frac{10^{29}}{d_\gamma}\right)^{1/2}
  \left(\frac{10^{12}\,{\rm G}}{B_p}\right) .
  \label{eq:NS-surface-penetration}
\end{equation}
For the wind region, one finds instead
\begin{equation}
  \begin{aligned}
    L_{\rm skin}^{(\gamma)}(r)
    \simeq{}&
    43.5\,{\rm km}
    \left(\frac{1.23}{\kappa_\Phi}\right)
    \left(\frac{10^{28}}{d_\gamma}\right)^{1/2}
    \left(\frac{10^{12}\,{\rm G}}{B_p}\right) \\
    &\times
    \left(\frac{P}{0.1\,{\rm s}}\right)^3
    \left(\frac{10\,{\rm km}}{R_\star}\right)^3 
    \left(\frac{r}{R_{\rm LC}}\right)^2 .
  \end{aligned}
  \label{eq:wind_penetration}
\end{equation}
Equating \(L_{\rm skin}^{(\gamma)}(r)\) with \(r\) gives the radius beyond which
the wind no longer behaves as an opaque barrier,
\begin{equation}
  \begin{aligned}
    r_{\rm Born}
    \simeq{}&
    5.2\times10^5\,{\rm km}
    \left(\frac{\kappa_\Phi}{1.23}\right)
    \left(\frac{d_\gamma}{10^{28}}\right)^{1/2}
    \left(\frac{B_p}{10^{12}\,{\rm G}}\right) \\
    &\times
    \left(\frac{R_\star}{10\,{\rm km}}\right)^3
    \left(\frac{0.1\,{\rm s}}{P}\right) .
  \end{aligned}
  \label{eq:NS_rBorn}
\end{equation}
For the same benchmark pulsar with radius $R_\star = 10\,$km, 
surface magnetic field strength $B_p = 10^{12}\,$G and
$\kappa_\Phi=1.23$,
$r_{\rm Born}\simeq 5.2\times10^5\,{\rm km}$ for $d_\gamma=10^{28}$, 
and $r_{\rm Born}\simeq 5.2\times10^3\,{\rm km}$ for $d_\gamma=10^{24}$.

The neutron star contribution depends on their magnetospheric cross sections and more concentrated density profile. 
However, their low abundance in the adopted population makes their contribution to cloud depletion negligible compared with that of ordinary stars.

\paragraph{Implications for cloud growth.}

\begin{figure*}[!t]
    \centering
    \includegraphics[width=\textwidth]{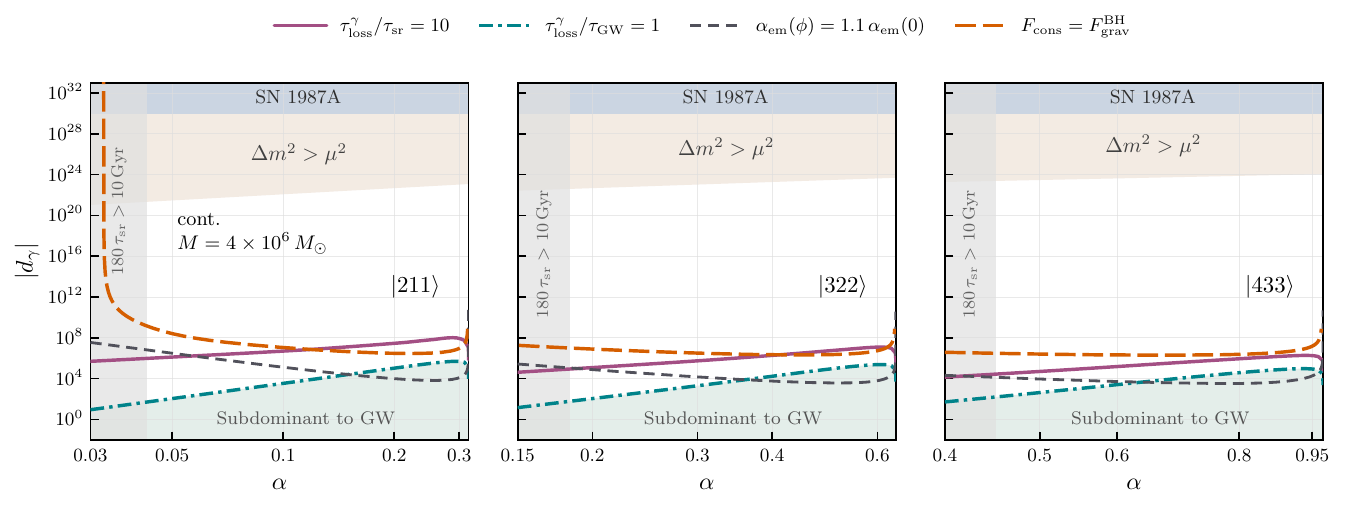}
    \caption{
    Parameter space for the quadratic scalar–electromagnetic coupling with
    $d_\gamma < 0$, 
    using the same BH and stellar
    population as in Fig.~\ref{fig:dN_alpha_mw}.
    The purple, green and orange curves follow from the matching in Eq.~\eqref{eq:photon_nucleon_conversion}.
    The blue-gray shaded region instead shows the independent
    photon coupling SN1987A cooling bound, $|d_\gamma| < 9.2 \times 10^{29}$~\cite{Gan:2025nlu}.
    The black dashed curve corresponds to a maximum increase of 10\% in $\alpha_{\mathrm{em}}$, 
    evaluated at the point of highest cloud density, 
    with the cloud at its saturation mass for each state and value of $\alpha$.
    } 
    \label{fig:dgamma_alpha_mw}
\end{figure*}

The energy loss calculation proceeds in the same way as in the quadratic nucleon coupling case. 
We use the same orbit-averaged loss formula in Eq.~\eqref{eq:cloudmasslossEccen}, 
replacing the electromagnetic coupling induced scattering length $a_s$ -- this gives the cloud depletion time $\tau_{\rm loss}^\gamma$. 
The corresponding cross sections are shown in Fig.~\ref{fig:stellar_density_profiles} for ordinary stars at different values of $d_\gamma$ and in Fig.~\ref{fig:momentum_transfer_cs} for neutron stars at different surface magnetic field strengths $B_p$. 
They are discussed further in Appendix~\ref{app:StellarProfile}.

For ordinary stars, Eq.~\eqref{eq:photon_nucleon_conversion} gives
$d_N^{\rm eff}\simeq-1.3\times10^{-3}d_\gamma$. 
For the adopted elemental composition and negligible macroscopic magnetic fields, 
this substitution yields the same scattering length and cloud depletion rate when the same stellar profiles and rate prescription are used,
$\Gamma_\gamma(d_\gamma)\simeq\Gamma_N(d_N = -1.3\times10^{-3}d_\gamma)$.
The corresponding stellar cross sections are discussed in Appendix~\ref{app:StellarProfile}.
Figure~\ref{fig:dgamma_alpha_mw} summarizes the effects on cloud growth and decay for $d_\gamma < 0$. 
The purple solid curve marks the cloud depletion rate equal to 10\% of the superradiant growth rate, 
while the green dash-dotted curve marks equality with GW losses at saturation.
These curves are obtained by mapping the nucleon results in Fig.~\ref{fig:dN_alpha_mw} using the relation outlined above for ordinary stars.
Above these respective curves, stellar scattering appreciably slows cloud growth or shortens the cloud lifetime. 
We also use the same diffuse matter density, 
$\rho_N=10^{-18}\,{\rm g\,cm^{-3}}$, 
to identify the light brown shaded region, where $\Delta m_\gamma^2>\mu^2$. 
In addition to these mapped results, 
the blue gray shaded region shows the parameter space excluded by the independent SN1987A bound, $|d_\gamma|\lesssim 9.2\times 10^{29}$, derived from scalar pair production through $\gamma \gamma \rightarrow \phi \phi$
\footnote{The Primakoff-like process $\gamma p \rightarrow p \phi\phi$ provides an additional SN cooling channel, where $p$ denotes the proton.
But its contribution is a factor of 23 smaller than that of $\gamma\gamma\rightarrow \phi\phi$ in the proto-neutron star environment~\cite{Cox:2024oeb}. 
We can therefore neglect its contribution when deriving the bound.}~\cite{Cox:2024oeb,Gan:2025nlu}.
The black dashed curve corresponds to a 10\% increase in the fine-structure constant induced by the scalar background,
\begin{equation}
\frac{\alpha_{\rm em}(\phi)}
{\alpha_{\rm em}(0)}
=
\left[1-\frac{|d_\gamma|\phi^2}
{M_{\rm pl}^2}\right]^{-1}
=
1.1.
\end{equation}
For each state and value of $\alpha$, 
we take the cloud profile to its saturation mass and evaluate the field at the point of maximum cloud density.
Above this curve, the shift in
the fine structure constant exceeds 10\% if the cloud reaches its saturation mass. Note that for order one shifts the fine-structure constant actually becomes negative, and thus the 10$\%$ threshold should be thought of as an indicative point where the theory becomes problematic.
A smaller cloud mass, however, reduces
this correction and shifts the curve upward at fixed $\alpha$ and cloud shape, 
while leaving the cloud depletion rate,
and hence the purple curve, unchanged.
The black dashed curve therefore serves as a reference for a saturated cloud rather than a general constraint on the parameter space.
As in Fig.~\ref{fig:dN_alpha_mw}, we show the point at which the conservative force on a solar-mass star matches that of gravity (evaluated at the characteristic radius of the cloud) for a fully saturated cloud. 
For smaller cloud masses, this curve shifts to larger couplings, while the purple curve (comparing energy losses to superradiant injection) remains unaltered.



In the s-wave limit, 
the electromagnetic interaction also gives $a_s\simeq R_\star$, 
so Fig.~\ref{fig:quartic_ratio_mw_swave} describes the electromagnetic loss time as well. 
The continuous stellar-mass distribution therefore produces stronger depletion than the single solar mass cusp.
The same correspondence applies to the BH mass dependence in Fig.~\ref{fig:quartic_ratio_mass_scan_swave}, including the suppression of stellar depletion by tidal disruption at small BH masses.
At finite coupling,
the electromagnetic rate instead follows from
the nucleon calculation at the matched value of $d_N$ in Eq.~\eqref{eq:photon_nucleon_conversion}.

We find the same conclusion as in the quadratic nucleon-coupling case. 
For the MW case, the electromagnetic coupling can give $\tau_{\rm loss}^{\gamma}/\tau_{\rm sr}<1$ over a broad range of $\alpha$ for both stellar-population benchmarks. 
This means that stellar losses exceed superradiant growth and prevent net cloud growth. 
At the representative values of $\alpha$ shown, 
both stellar profiles also yield $\tau_{\rm loss}^{\gamma}/\tau_{\rm sr}<1$ over extended ranges of BH mass.

\section{Conclusions}
\label{sec:conclusions}

In this work we have investigated how the stellar environments around intermediate and supermassive black holes can influence the growth and evolution of superradiant scalar fields. In particular, we have demonstrated that: (1) purely gravitational dynamics are typically not expected to prevent the growth of superradiant clouds (although, slight modifications to the astrophysical modeling, or the inclusion of e.g.\ scalar self-interactions, could change this conclusion over limited regimes of parameter space), (2) purely gravitational dynamics are expected to change the lifetime of superradiant clouds, (3) a quadratic coupling of the scalar to nucleons or electromagnetism can suppress the growth of the cloud and enhance the decay rate of the cloud over large regions of parameter space.

For the three cloud state benchmarks considered in this paper, 
quadratic nucleon couplings leave the superradiant growth rate largely unaffected when $d_N\lesssim9\times10^2$ for the single-solar-mass cusp or $d_N \lesssim 50$ for the strongly segregated continuous stellar mass distribution. 
The corresponding bounds on the quadratic electromagnetic coupling follow from $d_N \simeq -1.3\times10^{-3} d_\gamma$.
Cloud growth and depletion depend sensitively on the masses, abundances, and distributions of stars and compact remnants. 
Predictions for other galactic nuclei therefore require a better understanding of their stellar and compact remnant populations.



\section*{Acknowledgements} YC is supported by the National Research Foundation of Korea (NRF) Grant RS-2023-00211732, 
by the Samsung Science and Technology Foundation under Project Number SSTF-BA2302-05, by the POSCO Science Fellowship of POSCO TJ Park Foundation, and by the NRF Grant RS-2024-00405629. AC is supported by an ERC STG grant (“AstroDarkLS”, grant No. 101117510). AC acknowledges hospitality by the Institute of Advanced Study supported by an anonymous donor. G.M.T.\ gratefully acknowledges support from the Rubicon Fellowship, awarded by the Netherlands Organisation for Scientific Research (NWO), grant \href{https://doi.org/10.61686/WYKDB06497}{https://doi.org/10.61686/WYKDB06497}.
SJW acknowledges support from a Royal Society
University Research Fellowship (URF-R1-231065). This
work is also supported by the Deutsche Forschungsgemeinschaft under Germany’s Excellence Strategy—EXC
2121 ``Quantum Universe"—390833306. 
\appendix

\section{Stellar profiles and scattering cross sections}
\label{app:StellarProfile}

The matching between an extended target and the
pointlike interaction used in the ionization calculation
is encoded in its low-energy transport cross section $\sigma_T$. For
the repulsive interactions considered here, the scattering
length is $a_s = \sqrt{\sigma_T(0)/4 \pi}$.

\begin{figure*}[t]
    \centering
    \includegraphics[width=0.48\linewidth]{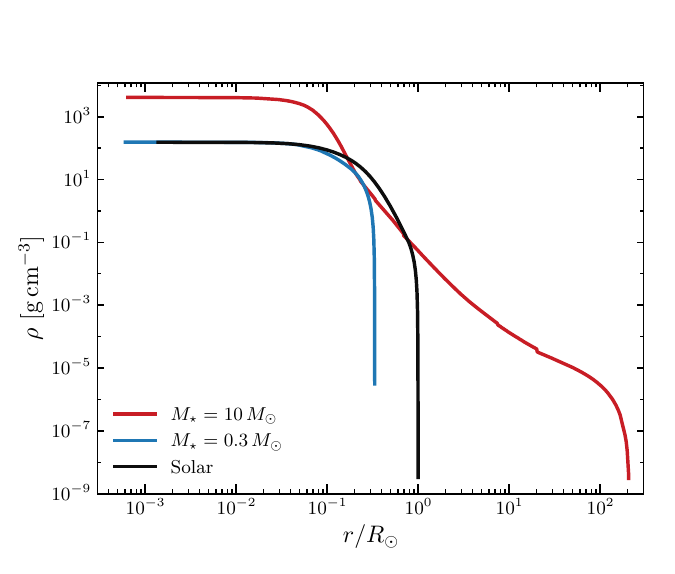}
    \hfill
    \includegraphics[width=0.48\linewidth]{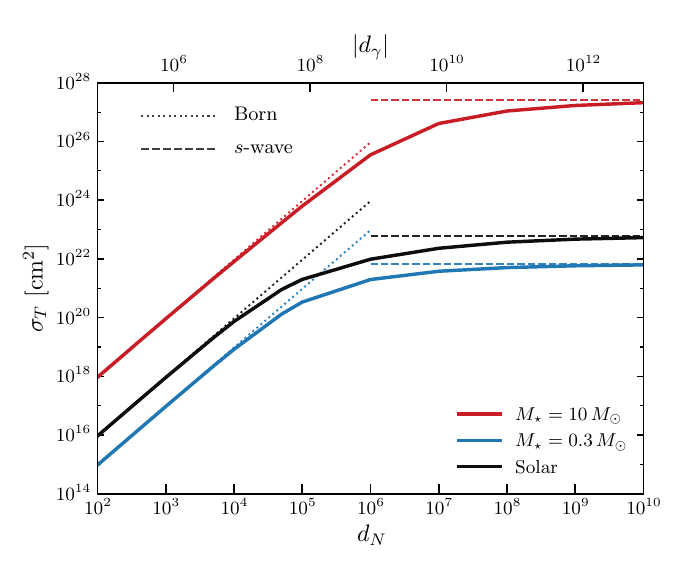}
    \caption{
    Left panel: stellar density profiles $\rho$ for stars of different masses, computed using \texttt{MESA}.
    Here we assume that the star with $M_\star = 0.3\,M_\odot$ is in the mid-main-sequence stage,
    while the star with $M_\star = 10\,M_\odot$ has already evolved into the red-supergiant stage.
    Right panel: the corresponding  momentum-transfer cross sections $\sigma_T$.
    The lower axis shows $d_N > 0$ while the upper axis shows the corresponding $|d_\gamma|$ for $d_\gamma < 0$.
    The long- and short-dashed lines correspond to the  $\sigma_T$
    obtained from the $s$-wave and Born approximations, respectively.
    }
    \label{fig:stellar_density_profiles}
\end{figure*}

Left panel of Fig.~\ref{fig:stellar_density_profiles} shows the density profiles and transport cross
sections for the $M_\star = 0.3 M_\odot$, solar, and $M_\star = 10 M_\odot$ benchmarks.
The first two are main sequence stars, whereas
the star with $M_\star = 10 M_\odot$ is assumed to be a red supergiant. 
In the right panel, the lower axis gives $d_N > 0$, 
while the upper axis shows the corresponding $|d_\gamma|$ for the repulsive case.
Their values are related by $d_N=-1.3\times10^{-3}d_\gamma=1.3\times10^{-3}|d_\gamma|$.
At small $d_N$, the numerical cross sections follow the Born prediction and scale as $d_N^2$. As  the coupling increases, the penetration depth decreases and the numerical result of $\sigma_T$ approaches the s-wave value, $\sigma_T \simeq 4 \pi R_\star^2$.
The different plateaus therefore reflect the stellar radii, 
while the transition between the two regimes also depends on the full density profile.

\begin{figure}[!t]
    \centering
    \includegraphics[width=0.468\textwidth]{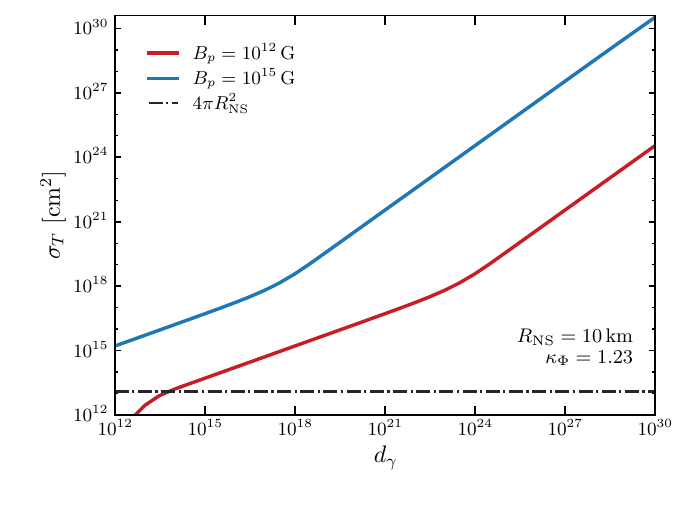}
    \caption{The corresponding  cross sections for neutron stars with different scalar electromagnetic couplings $d_\gamma$. 
    The red and blue lines correspond to neutron-star surface magnetic fields of $10^{12}\,{\rm G}$ and $10^{15}\,{\rm G}$, respectively.
    The dash-dotted line marks the geometric surface value $4 \pi R^2_{\rm NS}$ for $R_{\rm NS} = 10 \,$km.
    } 
    \label{fig:momentum_transfer_cs}
\end{figure}

Figure~\ref{fig:momentum_transfer_cs} separately shows the cross sections for the repulsive magnetic neutron star benchmark, for which $d_\gamma > 0$. 
Both the dipolar near zone and the open-field wind contribute to scattering. 
Increasing $d_\gamma$ or the surface field strength increases the radius out to which the magnetosphere is effectively opaque to the scalar field, 
allowing the cross section to exceed the geometric limit, $4 \pi R^2_{\rm NS}$. 
This enhancement represents scattering from the extended magnetosphere rather than from the neutron star surface.

\section{Comparison with gravitational wave depletion time $\tau_{\rm GW}$}
\label{app:GWdepletion}

A saturated cloud can also be slowly depleted through GW emission. 
The corresponding mass loss rate can be written as
\begin{equation}
    \dot M_c
    =-\frac{\gamma_j^{\rm GW}}{\alpha M}M_c^2\,,
    \label{eq:GWdepletion}
\end{equation}
where $j$ labels the cloud state and $\gamma_j^{\rm GW}$ is the
coefficient governing the depletion of the total cloud mass.
Using the leading flat space analytic results of Ref.~\cite{Yoshino:2013ofa}, we have
\begin{align}
    \gamma_{\ket{211}}^{\rm GW}
    &\simeq 1.56\times10^{-3}\,\alpha^{14}\mu\,,\\
    \gamma_{\ket{322}}^{\rm GW}
    &\simeq 1.89\times10^{-7}\,\alpha^{18}\mu\,,\\
    \gamma_{\ket{433}}^{\rm GW}
    &\simeq 2.89\times10^{-11}\,\alpha^{22}\mu\,.
    \label{eq:appdepletionTGW}
\end{align}
These rates are much slower than the growth timescales of the same corresponding state (but faster than the growth of the next state); nevertheless, the decay timescale plays an important role in establishing how long a fully grown superradiant cloud remains present in the system, and thus enters as a key ingredient in various searches for superradiance.

Equation~\eqref{eq:GWdepletion} has the solution
\begin{equation}
\begin{aligned}
    M_c(t)
    &=\frac{M_{c,0}}{1+t/\tau_j^{\rm GW}}\,,\\
    \tau_j^{\rm GW}
    &\equiv
    \frac{\alpha M}{\gamma_j^{\rm GW}M_{c,0}}
    =\frac{\alpha}
    {\gamma_j^{\rm GW}\left(M_{c,0}/M\right)}\,.
\end{aligned}
\end{equation}
Here $M_{c,0}\equiv M_c(0)$ is the initial cloud mass. By construction,
$\tau_j^{\rm GW}$ is the time required for the cloud mass to decrease
to one half of its initial value.

\begin{figure*}[!t]
    \centering
    \includegraphics[width=\textwidth]{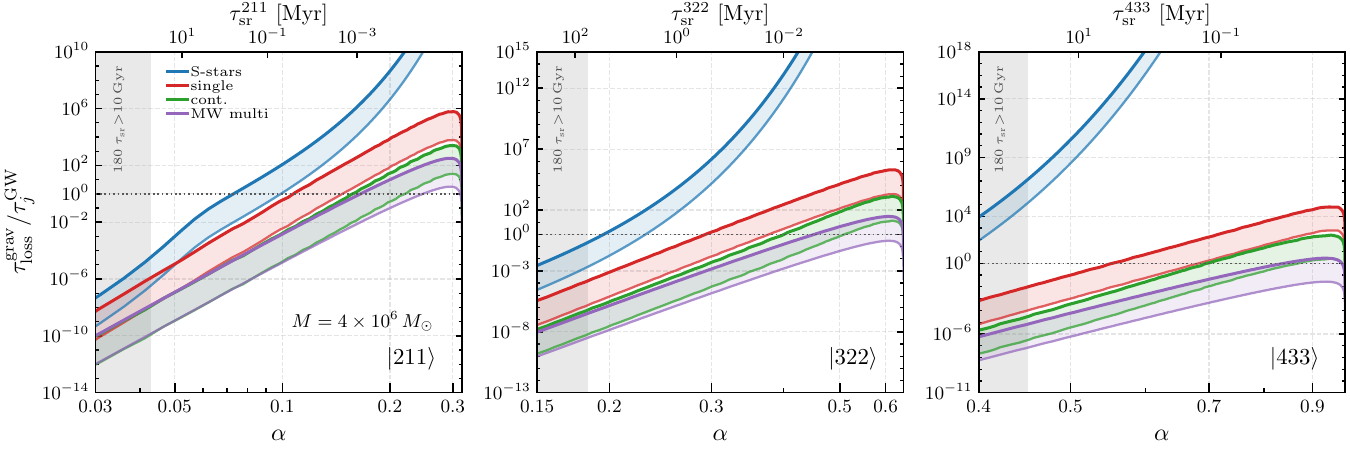}
    \caption{
    Ratio of the gravitational dynamical friction depletion time to the gravitational wave annihilation depletion time, $\tau^{\rm grav}_{\rm loss}/\tau^{\rm GW}$, for Sag A$^*$ with $M = 4 \times 10^6 M_\odot$.
    The three panels correspond to the $|211\rangle$, 
    $|322\rangle$ and $|433\rangle$ cloud states. 
    Blue, red, green and purple curves correspond to the S-star population, the single solar-mass cusp, the strongly segregated multi-mass cusp, and the  Milky Way multi-population profile, respectively. 
    The shaded bands show the range obtained by taking the cloud mass fraction between 
    $q_{\rm sat}(\alpha)$ and $10^{-2}q_{\rm sat}(\alpha)$, 
    where $q_{\rm sat}$ is the single-level spin-extraction saturation estimate for initial spin $\tilde{a}_i=0.9$.
    The low-$\alpha$ gray region indicates where 180$\tau_{\rm sr}$ exceeds the age of the  Universe.
    The dotted horizontal line marks $\tau^{\rm grav}_{\rm loss}=\tau^{\rm GW}$.
    } 
    \label{fig:gravity_gw_ratio_mw}
\end{figure*}

\begin{figure*}[!t]
    \centering
    \includegraphics[width=\textwidth]{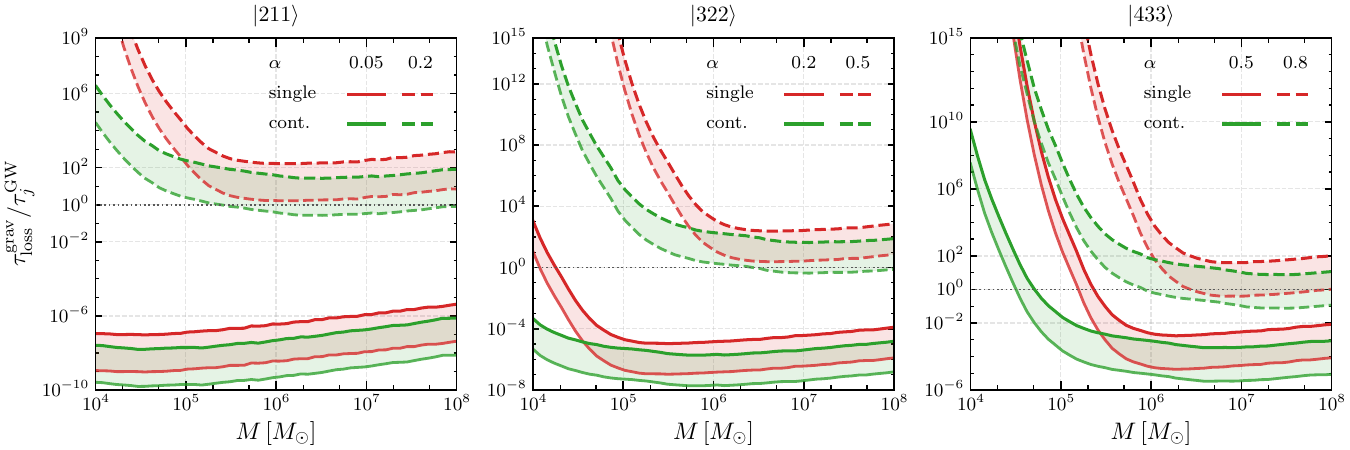}
    \caption{
    Ratio of gravitational dynamical friction depletion time to the gravitational wave annihilation depletion time in the BH-mass scan. 
    The three panels correspond to the  $\tau^{\rm grav}_{\rm loss}/\tau^{\rm GW}$ for the $|211\rangle$, $|322\rangle$, and $|433\rangle$ states.
    The stellar profiles are the single solar-mass cusp and the strongly segregated multi-mass cusp. 
    For each curve, the shaded band corresponds to varying the cloud mass fraction between 
    $q=q_{\rm sat}$ and $q=10^{-2}q_{\rm sat}$, where $q_{\rm sat}$ is the single-level spin-extraction saturation estimate for initial spin $\tilde{a}_i=0.9$. 
    }   
    \label{fig:gravity_GW_ratio_mass_scan}
\end{figure*}

Fig.~\ref{fig:gravity_gw_ratio_mw} and ~\ref{fig:gravity_GW_ratio_mass_scan} compare the gravitational dynamical-friction depletion time, $\tau^{\rm grav}_{\rm loss}$, with the gravitational wave depletion time, $\tau^{\rm GW}_j$, by varying the initial cloud mass $M_{c,0}/M$ from the saturation value of the $j$th state down to $10^{-2}$ of that value.
For the Milky Way benchmark, we find that $\tau_{\rm loss}^{\rm grav}/\tau_j^{\rm GW}<1$ over a wide range of $\alpha$, 
indicating that gravitational dynamical friction depletes a fully grown cloud more rapidly than gravitational wave emission.

The dependence on the BH mass is shown in Fig.~\ref{fig:gravity_GW_ratio_mass_scan}. 
For each cloud state, 
stellar depletion generally dominates at smaller values of $\alpha$. 
At larger values of $\alpha$, however, the cloud becomes more compact, 
and for sufficiently small BH masses the stellar depletion is strongly suppressed because of the reduced cloud size. 
Combined with the steep scaling of the gravitational wave timescale in Eq.~\eqref{eq:appdepletionTGW}, 
gravitational wave emission becomes the dominant depletion mechanism. Note that for quadratic interactions, there is also a broad range of parameter space where the dissipation timescale can also be much shorter than the gravitational decay timescale. As shown in Fig.~\ref{fig:dN_alpha_mw} and Fig.~\ref{fig:dgamma_alpha_mw}, this occurs roughly for couplings to nucleons or electromagnetism in the range 
$d_N \lesssim 1$--$8$ for the single-mass population and $d_N \lesssim 0.1$--$0.5$ for the continuous population,
with each range spanning the three states for saturation cloud mass.
The corresponding electromagnetic couplings are $|d_\gamma| \simeq (1.0 - 5.8) \times 10^3$ and $|d_\gamma| \simeq (1.0 -3.4) \times 10^2$, respectively,
on the repulsive branch with $d_\gamma < 0$.

\section{Comparison on ionization and local dynamical friction}\label{app:ComparisonIDF}

\begin{figure*}[t]
    \centering
    \includegraphics[width=0.48\linewidth]{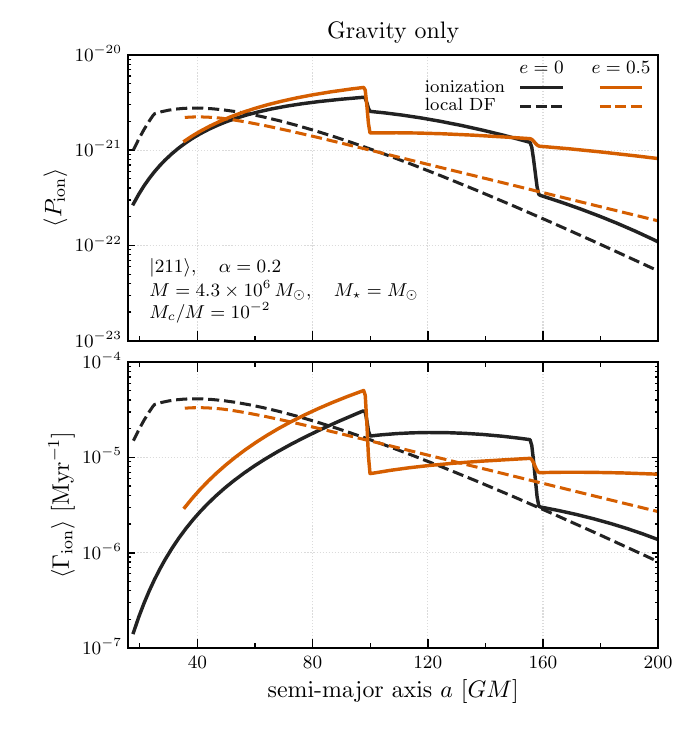}
    \hfill
    \includegraphics[width=0.48\linewidth]{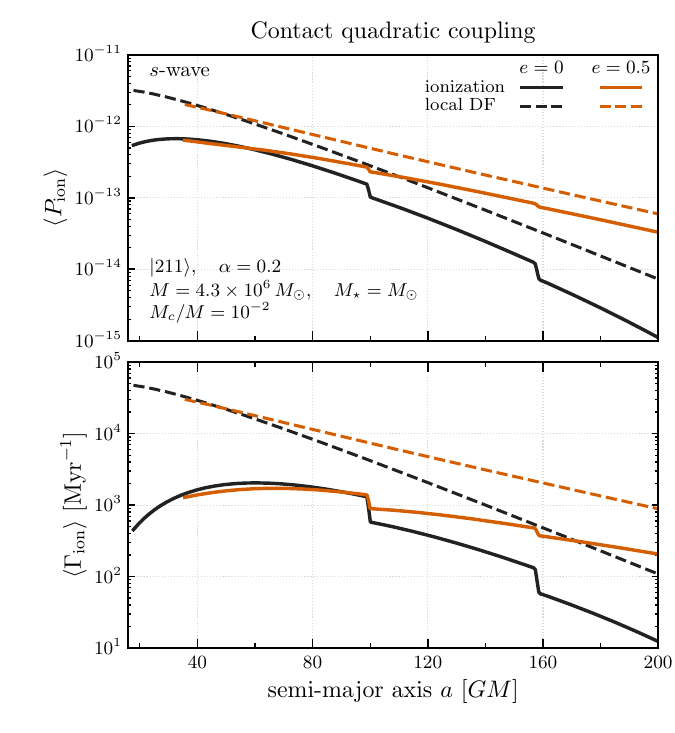}
    \caption{
    Comparison between the ionization treatment and the local dynamical friction approximation for the gravitational interaction (left) and the quadratic
    contact interaction (right).
    In each case, the upper panel shows the orbit-averaged powers
    $\langle P_{\rm ion}\rangle$ (solid) and
    $\langle P_{\rm DF}\rangle$ (dashed), 
    while the lower panel shows the
    corresponding cloud depletion rates
    $\langle\Gamma_{\rm ion}\rangle$ and
    $\langle\Gamma_{\rm DF}\rangle$.
    Black and orange curves correspond to the orbits with eccentricity $e=0$ and $e=0.5$,
    respectively.
    We take the BH mass $M=4.3\times10^6\,M_\odot$ and the solar like star as benchmark.
    The values $\alpha=0.2$, $q = M_c/M=10^{-2}$ are chosen for the
    $\ket{211}$ state. 
    Each curve begins where the orbital pericenter equals the tidal disruption radius $r_t = a (1-e)$.
    }
    \label{fig:comparsion_DF_ion}
\end{figure*}

Figure~\ref{fig:comparsion_DF_ion} shows the explicit comparisons between the ionization treatment and the local dynamical friction approximation for both the gravitational and contact quadratic potential cases.
The upper panels compare the orbit averaged ionization power,  $\langle P_{\rm ion}\rangle$, obtained by explicitly summing the transitions from the bound cloud state to continuum states, with the corresponding averaged local dynamical friction power, 
$\langle P_{\rm DF}\rangle$, calculated using Eq.~\eqref{eq:dynamicalFGR}  (gravity only)  and Eq.~\eqref{eq:s-waveSigma} for the case of s-wave scattering.
The lower panels compare the direct ionization rate, 
$\langle\Gamma_{\rm ion}\rangle$, defined in Eq.~\eqref{eq:ionmdot}, with the ionization rate inferred from local dynamical friction, 
assuming that the entire orbital energy is used to overcome the binding energy of the cloud,
\begin{equation}
    \left\langle\Gamma_{\rm DF}
    \right\rangle
    =
    \frac{\mu \left\langle P_{\rm DF}\right\rangle}{M_c |\epsilon_b|},
\end{equation}
where $\epsilon_b = -\alpha^2 \mu/(2 n_b^2)$ is the binding energy.
For the orbit with small semi-major axes, 
the two methods differ significantly, 
with the ionization calculation generally predicting a lower cloud depletion rate than
the naive local dynamical friction estimation. 
In this regime, the energy extracted from the stellar orbit is well above the ionization threshold, 
with a substantial fraction converted into the kinetic energy of the outgoing cloud particles. 
At larger semi-major axes, the available orbital energy lies closer to the binding threshold, leading to much closer agreement between the two methods.

\section{Energy losses driven by time-dependent background}
\label{app:ConservativeF}

In addition to the dissipative scattering studied in the main part of the text, the quadratic interactions generate a conservative force that stems from the diagonal matrix element of the perturbation. Namely, for a cloud with occupation number $N_b(t)$, the effective force felt by the star is given by $F_{\rm cons} = - N_b(t) \, \nabla \left<b | V | b \right>$, which yields a net energy transfer over an orbit given by
\begin{eqnarray}
    \delta E = - \oint F_{\rm cons} \cdot d \ell = -\delta\left[N_b \, \left<b | V | b \right>\right]\big|_{\rm orbit} + \nonumber \\ \oint dt \, \partial_t N_b(t) \, \left<b | V | b \right> \, .
\end{eqnarray}
Assuming the cloud is neither growing efficiently, nor dissipating efficiently, during an orbit, $\delta E \sim 0$ (which follows from the fact that the force is conservative). For some non-trivial temporal evolution, however, one can generate energy dissipation on the order of $\delta E \sim \delta N_b \times \left<b | V | b \right>$. We have computed this effect explicitly for a handful of systems of interest, and it was always found to be highly subdominant to the  bound-to-free transition calculation computed in the main text.

Finally, let us mention that we have also neglected the effect of having a temporal background arising from the rotation of the star itself -- e.g., a dipolar field will induce an inhomogeneous scattering surface that can impart energy on the order of the rotational frequency. We have verified that for the couplings and systems studied here, we do not expect this effect to be a sizable correction to the rate of energy dissipation.

\clearpage

\bibliography{biblio}

\end{document}